\documentclass[12pt,preprint]{aastex} 
\usepackage{color}                                                                   
\begin{document}

\title { A Model For Dark Matter Halos.}

\author{H. L.  Helfer}

\affil{  Dept. of Physics and Astronomy, The University of Rochester,}
\affil{ Rochester, NY 14627}

\begin{abstract}
      
      A  dark matter halo model is developed postulating a new state of matter,   entities which have  internal spin-like terms.  Their motion in an external Schwarzschild metric is discussed. The internal  spin motion contributes to the centrifugal force along with the usual orbital angular momentum term and can severely limit the distance of closest approach to the attractor. An energy-momentum tensor associated with an aggregate of them is shown  to have  primarily pressure-like components. A model of the spiral galaxy  halos is developed which can match the observed `flat' rotation curves of some galaxies. The halo  dark matter `missing mass' results from the  pressure term's contribution to the metric tensor.   An addendum to the standard cosmology picture allows an estimate of the amount of dark matter; this is in reasonable agreement to that observed.  It is possible that the adopted representation of the internal spin motion could be replaced by a boson string Lagrangian.
       
        \end{abstract}

         {\bf   \ \ \ \ \ \ \ \ \ \ \ \ \ \ \ \ \ \     Preliminary remarks: Observational Constraints  } 
               
       There exist many spiral galaxies (including the Milky Way) with `dark matter' halos.  The observational characteristic of spiral galaxy halos are galaxy rotation curves, $v_{rot}(r)$ consisting of: (1) a central spherical contribution followed by a (roughly) linear part :    $v_{rot}(r) = V_0(r/r_{00})$ for $r \le r_{00} \sim 2\ -\ 4$ kpc;  (2) a (fairly) `smooth' transition zone $r_{00} \leq r \leq r_0 \sim 4- 8$ kpc;  and (3) a ``flat" part $v_{rot}(r) = V_0$ for $r_0 < r \le r_1 \sim 16\ - \ 50+$ kpc. This outer limit is  hard to estimate and in a few cases may be $ \geq 100 $ kpc. A  value $V_0 \sim 200\ {\rm km \ s}^{-1}$ characterizes large spirals. [1,2,3,4,5,6]
       
       These rotation velocities\footnote{These are defined by the observed  line-of-sight motions of extreme   Population I objects such as very luminous HII regions or  HI gas and molecular clouds, known to depart by less than $\sim 5 \%$ from  circular orbital velocities $v_{cir}(r)$ in galactic disks outside the very central region. The absence of these objects in the outer parts of galaxies limits our present knowledge of the rotation curves at large $r$. } are interpreted as circular velocities $v_c$, with $v_c^2/r=-\ \partial \Psi/\partial r$ implying that the the potential is given by
       $$\Psi_{obs} \cong k- v_c^2 \ln (r/r_0)\ {\rm for}\   r_1 \geq r \geq r_0. $$
       Alternately, one sets $v_{cir}^2 =G{\cal M}_{obs}/r$  where ${\cal M}_{obs} \propto r$ when $ r\geq r_0$ and regards  ${\cal M}_{obs}$ as a true mass distribution.  No galactic mass distribution compatible with the observed stellar distributions predicts this behavior in the outer zone.[15]  An unseen (`dark') matter component is  usually hypothesized, probably amounting to several times the mass contribution  inferred for the stellar and interstellar matter contribution. There is no  evidence  that  this `dark' matter (DM)  interacts with ordinary matter (OM) through electromagnetic interactions or collisions.  Because the  halos are concentrated, the   halo DM must be moving at non-relativistic velocities 
       $v < \ \sim 10^3 - 10^4\ {\rm km \ s}^{-1}$. 
                                                    %pg 2
      
        One may also infer that it is not present in the stars of the Milky Way galaxies (and other galaxies).   This is an important observation. Theoretical stellar models developed in the  last fifty years  to represent the relationships between masses, radii, luminosities  and ages  of field and cluster stars have been quite successful.  These   basically require as input   specification of the  opacity per gram and the energy generation per gram.   The presence of a significant inert DM component  contributing to stellar densities would cause  these models to fail to represent observations.  An upper limit of the  DM contribution to stellar models is probably about the uncertainty in the He abundances; perhaps $ \sim 5 \%$.   We infer that it is difficult, if not impossible, to bind DM in  stellar gravitational wells.
       
        There have been many DM proposals [7], including explaining the rotation curve results by modifying Newtonian dynamics [8] or by introducing a new scalar field [9,25].   The rationale for introducing yet another model is to additionally explain the  absence   of DM in stars, to provide a connection to the DM used in the standard cosmological models (where it is needed to  produce satisfactory nucleosynthesis results [10,11]) and to raise the possibility of the existence  of a radically different new state of matter.  
        
         We note that the flat rotation curves are  completely equivalent to the expression for one of the Schwarzschild metric coefficients, Eqn (C3), with an unusual source term.  So the focus is on trying to model this source term.   This new state consists of entities which can carry momentum transverse to the  usual (time-like) four-momentum characterizing ordinary particles.   Using aggregates of such entities,  one can explain both the rotation-curve observations and the absence of DM in the interior regions of galaxies.  The parameters used for fitting the observed rotation curves  are interpreted as describing properties of  galaxies moving through a local DM intergalactic medium.              
                       This paper is divided into three parts.   To infer properties of DM from the observations, I first use a rather detailed  procedure for determining a form for the source term in GR, the energy momentum tensor $T^{\mu\nu}$.  Basically  one starts with the construction of the source term from the equation of motion for a stream  of matter and conservation of mass along it.  Averaging collections of  such OM streams  can give the standard $T^{\mu\nu}$ source term;  the lack of need for a thermodynamic equation of state is emphasized.  In the second part a proposal for DM streams, that they carry momentum  in a direction transverse to its average motion, is explored.  Their motion in a central force field is examined and it shown that  they possess an intrinsic angular momentum which prevents their close approach to the attractor. Finally two different forms for $T^{\mu\nu}$ for collections of DM streams are derived, one of which requires introduction or a term similar to  the  cosmological `constant'.  A model of a galactic halo is constructed using  the other form  which features a dominant `false' pressure term;  this model reproduces the observed `flat' rotation curves.

\subsubsubsection{ The Assumed Properties Of  Classical Streams.}                     
                  In the small region\footnote{For mathematical conventions and details about  how sums of streams and orthogonality of contiguous streams are defined  in ${\cal R}$ see the Appendices: A1-A4.}  ${\cal R}$  of the universe under study  there are paths $x^\mu(\tau)$ which are geodesics ( $ dx^\mu /d\tau  \equiv U^\nu U^\mu_{;\nu}=0$, defining the tangent vectors $U^\mu(\tau),$) that neither end nor begin inside the region.  We take as  a primitive  concept  a  classical isolated {\it  stream}.  It consists of
 objects traveling together along such  a path. They have the same  normalized (time-like) velocity $U$. The stream  is characterized by a mass density $m(x^\mu(\tau)) \not\equiv 0$ (and a  very small but finite cross-section, $\Delta^3x$, which need  not be explicitly displayed and can be thought of as constant. 
                  
                  When needed  the stream's internal  structure will be represented by a local velocity field, limited by the cross-section, in which the velocities are parallel to the tangent of the central geodesic.  The normalization of  each streamline's velocity would reflect the relative density distribution across the stream.  We assume that this local field need not be specified  except when interactions with other streams are considered.  
           A stream  is used here to represent only  kinematic properties of an aggregate.  It  is intended to be a limited classical analog of the quantum mechanical $\langle p\mid p\rangle$, the density distribution of a particular momentum state. We  shall distinguish one stream from another by explicitly adding the subscript label $s$; the  label $s$ contains all the information needed to identify a particular stream.                                                                                                                                                                                                                                                           Primitive streams are more restricted than  vector fields; classically they do not ``add" unless they have a point in common. (However, in appendix A3 the concept of `addition ' is expanded to include contiguous streams in forming local {\it field} averages.)   
                               
  \section{  The Representation Of The Energy-Momentum Tensor  By Streams.} 
             {\it  ``...about the dread right-hand side of the Einstein's equations..." }[11]                                                                             
                                                                                                                                   
                                                                                                 %pg 3.

      For each stream $s$ we define an energy momentum tensor 
      $$T_s^{\mu\nu} \equiv m_s U^\mu_sU^\nu_s$$
       along the path and determine the variation of $m_s$ by requiring
            $$ (m_sU^\nu)_{;\nu} \equiv 0, \ \ \ {\rm or}      \eqno (1.1a)$$
            $$d m_s/d\tau_s + m_s U^\nu_{s;\  \nu} =0, \eqno(1.1b)$$
then  $T^{\mu\nu}_{s \ ;\nu}  = 0$ and $T^{\mu\nu}_s $ can be used as a source term for determining  the curvature tensor. 

     One may show (see the Appendix A5)  that $U^\nu_{s;\  \nu} =0$. Then the equations have three useful forms depending upon the functional form of $m_s$ available: (1) steady state one-dimensional pipe flow: 
$$m_s v_s ={\rm constant};$$                                                                                                    
(2)  given the {\it fluid} form $m_s(\tau)=m_s( x(\tau))$, the  equation of mass continuity results,     
 $$ \partial_t \ m +   {\vec v} \cdot \nabla m=0,$$
 where $U^\mu_s =\Gamma (1,{\vec v})$; and (3), given the {\it kinetic} form $m_s(\tau)=m(\ x(\tau), U^\mu(x(\tau))\ )$, Liouville's equation results, 
 
  $$[U^\mu\partial_\mu -\Gamma^\mu_{\alpha\beta }U^\alpha U^\beta (\partial/\partial U^\mu] m=0,$$
 which plays an extremely important role in interpreting stellar kinematics.[12,13]  Each of these equations hold only on a time-like path $x^\mu(\tau)$.                              
 
 \subsubsection{Ensembles Of Streams}                             %pg 4
 
 Suppose we have a finite number of streams in a small  region ${\cal R}$.  The conventional procedure is to treat  ${\cal R}$  as an energy-momentum reservoir and effectively define local {\it fields }  $\bar N$, $\bar U$ and $ T^{\mu\nu}$ extending throughout  ${\cal R}$ as averaged values of the included streams.  This averaging process is really non-trivial and as a consequence, we shall argue that the assumption that locally a thermodynamic   `equation-of-state'  is needed is moot.

           To introduce  fields as mass-averaged  quantities  representing  sums of streams  in 
                ${\cal R}$, put $ U_s^\mu = {\bar U}^\mu  + \delta  U_s^\mu$, where
                                                                                                                                                               
       $${\bar N} {\bar U}^\mu = \Sigma_s m_sU_s^\mu,\ \ {\rm  with}\ \ {\bar N} =\Sigma_s m_s\ {\rm and}\  \Sigma_s m_s \delta  U_s^\mu =0.  \eqno (2)$$
               Here, ${\bar U}$ and ${\bar N}\equiv \surd (-g) \rho$ are  fields  characterizing the region ${\cal R}$, with $( {\bar N} {\bar U}^\nu )_{; \nu}=0$, giving the ensemble's energy density conservation,  and 
                                                                                                                                             
                 $$   T^{\mu\nu} \equiv \Sigma_s  T^{\mu\nu} _s  = F^{\mu\nu} + P^{\mu\nu},\ \ {\rm where }          \eqno (3a) $$
         $$    F^{\mu\nu} = {\bar N}{\bar U}^\mu{\bar U}^\nu \ \ {\rm and} \ P^{ \mu\nu} = \Sigma_s  m_s\  \delta  U_s^\mu \delta  U_s^\nu \eqno (3b)$$
         with $ T^{\mu\nu}_{\ \ ;\nu} =0$, so that $T^{\mu\nu}$ can be used as a source (energy-momentum) tensor.  The trace is $ T = g_{\mu\nu}T^{\mu\nu}  ={\bar N}$, the rest-mass density.                                                     %pg 5

                Ordinary matter is characterized  by constant  momenta locally, $U^\mu_{s;\xi} =0$ (allowing for coordinate transformations). Now, it is quite remarkable that
                                                                                          
         $$   P^{\mu\nu}_{\ \ ;\xi} =  \Sigma_s   \delta  U_s^\mu \delta  U_s^\nu\  m_{s,\xi} \ .   \eqno(4)$$          
         
      This follows by differentiation by parts and noting  that because $U^\mu_{s;\xi} =0$, one has $(\delta  U_s^\mu)_{;\xi} = -{\bar U}^\mu_{;\xi}$\ .   This means that in small regions the change of $ P^{\mu\nu}$ with position is independent of the affine connections, both frame forces and gravitational forces, the mean velocity ${\bar U}^\mu$, and of details of the variations of  each $\delta  U_s^\mu$ with position,  assuming the fluid form of $m_s$.  It is a function only of the stream densities  which can be quite variable, and possible non-kinetic properties. The tensor $P^{\mu\nu}$ specifies the variances of the streams' motions; these variances depend upon the densities for each species of streams present.  There is no reason to assume these relations are thermodynamic since the selection of streams for the averaging is not specified.

      If the summation  and averaging in Eqn(3b) is  actually performed then under coordinate transformations  $P^{ \mu\nu}$ must transform  as a tensor of the form $h^{\mu\nu}(\rho),\   h^\mu(\rho)h^\nu(\rho)\ {\rm or}\ h(\rho)g^{\mu\nu}$.  The lack of dependance on the affine connections  means that only the third form, the conventional fluid pressure term, is acceptable when the fluid form of $m_s$ is used and isotropy is assumed.

      If the kinetic form of $m_s$ is used, then {\it e.g.}  $m_{s,\xi} \to \Gamma^\alpha_{\beta \xi}U^\beta_s(\partial m_s/\partial U^\alpha_s)$ is permissible. Then the other two generic forms  of $P^{\mu\nu}$ are not excluded. Indeed they need to be used in constructing rigorous kinetic models of galaxies since the various classes of stars exhibit tri-axial velocity distributions.[14,15]. The quantity $P^{\mu\nu}$ is now similar to a hydrodynamic stress tensor.  For non-relativistic fluids, $U^0_s \approx 1$; then the  conventional GR  approximation,$ T^{00}\cong {\bar N}; \ T^{ij}  \sim T^{00}/c^2 \cong 0$,  is not useful in many astronomical problems.

\subsubsubsection{ The Contentious Nature Of The `Equation Of State'.}                                                                       %pg 6                                                                      
      Normally we treat ensembles of  streams which are locally isotropic  and have the property that in its rest frame $T^{\mu\nu}$ is diagonal, $(A,B,B,B)$, with all four elements positive.  Then  one may always write formally $T^{\mu\nu} = (\rho + p)U^\mu U^\nu -p g^{\mu\nu}$ and regard the requirement $T^{\mu\nu}_{\ ; \nu}=0$, as an `equation of state'.  But by construction Eqns.(3) exactly define a source function with $T^{\mu\nu}_{\ \ ;\nu} =0$; no  thermodynamic interpretation  is needed.   If we choose a standard background model which locally admits integrals of motion expressing the constancy of quantities like energy $E_s$ or angular momentum $J^\mu_s$ along a stream then  the summations  in Eqn s.(2,3) can  be replaced by integrations over frequency distributions of the conserved quantities [{\it e.g.} $\Sigma_s \to \int dE_s \ f(E_s-\bar{E})$ ]  with  $T^{\mu\nu}_{\ \ ;\nu} =0$  preserved.  Again, no equation of state is needed.  
      
     The real problem is that  while the summations are sufficient, resulting in values  for all components of $P^{\mu\nu}$ in a particular coordinate system, they do not determine a generic form for its tensor representation.  If $P^{\mu\nu}$ is spatially isotropic, then  an empirical form  such as  $h(\rho)g^{\mu\nu}$ for  $P^{\mu\nu}$ can be adopted, provided constraint equations   are also introduced, $g^{\mu\nu}h(\rho)_{,\nu}  +F^{\mu\nu}_{\ ;\nu}=0$,  so that  $T^{\mu\nu}_{\ \ ;\nu} =0$ holds. If $F^{\mu\nu}$ is assumed appropriately simplified because of isotropy (or choice of Lorentz frame), these four constraints may be reduced to one, an ``equation-of-state".  If $P^{\mu\nu}$ is not spatially isotropic, the kinetic representation of  $m_s(\tau)$ is needed  and the four constraints  may not collapse into one;  the scalar representation of the pressure would be inappropriate.    In the stream representation, once the fluid form of  $m_s(\tau)$ is adopted and specified, the form of $h(\rho)$ is  specified  up to an additive constant by the differential constraint. 
     
     We emphasize that a `true' pressure term only arises when the tensor $P^{\mu\nu}$ is included; often in astronomical applications this tensor is   ignored  because $p/c^2 \approx 0$. Consequently, when the standard fluid source term is used in these cases, the constraint  $T^{\mu\nu}_{\ \ ;\nu} =0$ simply means conservation of mass  and nothing more.  
     
     We infer, then, that the introduction of a $T^{\mu\nu}$ for DM , even when written in standard fluid form, does not require a thermodynamic interpretation.   

 \subsubsubsection{Boundary Conditions.}                                                                                             
   The  generic form used  $T^{\mu\nu}=(\rho +p)U^\mu U^\nu -pg^{\mu\nu}$ is actually ambiguous; because it represents a solution of differential equations $T^{\mu\nu}_{\ ; \nu} =0 ,$ one may  always introduce a constant  $\lambda$ by adding $-\lambda g^{\mu\nu}$  to  $T{\mu\nu}$, allowing the substitutions 
   $p \to {\hat p} =p + \lambda,\ \ \rho \to {\hat \rho} =\rho -\lambda$.  So $\lambda$ must always be assigned; this modifies the physical interpretation of the $p -\rho$ relation.  We suggest that when this fluid form of $T^{\mu\nu} $ is introduced one should interpret   $T^{\mu\nu}_{\ \ ;\nu} =0$  as defining $\delta p$, given $\delta \rho$.  Properly, one may introduce a $\lambda_s$ term for each stream, if boundary conditions warrant it. Then for an  ensemble,  a $\langle \lambda_s \rangle $ term may be used, and this, depending on {\it e.g.} the streams' energy distribution, may not be a constant. Normally,  boundary conditions apply only to an ensemble; then  $\lambda$ is constant.

\section{ The Representation Of Dark Matter By Streams}                  %pg 7            
       We suggest that DM is a relic of a prior stage of the universe in which some  matter streams transported  space-like momenta and could travel along the space-like geodesics with tangent velocities  proportional to their momenta. Something happened. In the present universe these DM streams  now travel upon time-like geodesics, with $U^0>0$, but must carry the excess momentum along with them.  We examine a mechanism for doing so.
  
 So far we have dealt  only with the components of  a stream's momentum that  are aligned with the flow along a time-like geodesic.   There is no transport of additional momentum (such as carried by `eddies').   We will use the term `transverse momentum'   to refer to momentum components orthogonal to a stream's (time-like)  tangent vector that accompany the motion along the time-like stream.  By using `close' pairs of streams, we show such transport can be defined.  These will be our  candidates for DM when the total momentum associated with a pair is space-like. We shall eventually see that the transported transverse momenta   contribute to the source terms $T^{\mu\nu}$ the  equivalent of large pressure terms.

\subsection{ The Transport Of Transverse Momentum.}             %pg 7     

       While (field) vectors at a point can be added we cannot really do the same for streams, because two streams can `intersect'   at most only at isolated points and we cannot add vectors at different points.  But   two streams  may be contiguous in a very small region, $\Delta {\cal V}$, and we may: (1) regard momenta transport in this  region to be that of the sum of the two streams as if they did actually overlap; and  (2), define orthogonality  for non-intersecting streams (See  Appendix A5 for details.) With this understanding, we may talk of the `addition' of the momenta of contiguous streams.

           The total momenta of any stream $K$ shall be  represented  by a pair of close streams: (1) a time-like $x^\mu(\tau)$  stream   (with tangent  $U^\mu(\tau))$defining the direction of the usual momentum flow along a time-like path with $U^0>0$; and (2),  a  space-like   $x^\mu(\sigma)$ local stream with tangent  $S^\mu(\sigma)$ representing the direction of the transverse   momenta being transported, with $U\cdot S=0$; by `local' we mean $x^\mu(\sigma)$ has no physical significance outside of   $\Delta {\cal V}$.                                       
           
           As a guide, conservation of momentum for   $K$  at  a path point can be conventionally represented by              
       
  $$m(x) K^\mu = m(x)\cdot ({\bar a}U^\mu +{\bar b}S^\mu), \eqno(5)  $$
  
  where ${\bar a},{\bar b}$ are factors to allow us to adopt the  normalizations $K^2=\pm 1,0, \ U^2=1,\  S^2=-1$. We consider local Lorentz  frames for which $K^0,U^0 >0$. [For any such pair, we can choose a particular Lorentz frame in which $S^0=0$.] We focus on the case   when $K$ is space-like, $K^2=-1$   There is a one parameter  set of $(U,S)$ satisfying eqn (5)  but the ambiguity of choice is not serious  for determining $\bar a, \bar b$  when the $U-$motion is non-relativistic. (see  Appendix B1.)  For the halo DM choose as a representative  $K=\Gamma (k,0,0,1)$; then one has  ${\bar a}  \cong \Gamma k,\  {\bar b} \cong \Gamma$.  %pg 8
  
     A model using two-stream transport to replace the stream $K$, will  be the basis for representing DM. The significance of $K$ is that it enforces the conservation of 4-momentum given by Eqn(5)  for streams in a local neighborhood. From the concentration of DM in the halos of galaxies and in clusters of galaxies, we infer that DM exhibits sub-relativistic velocities there and we need not represent DM by space-like geodesics alone.  
  
       We define a unit of DM to be a pair of  $U$ and $S$ streams that are contiguous in some small region $\Delta {\cal V}$.  The pair follows the path described by the $U$ vector.  The $S-$vector path has no physical significance outside of $\Delta {\cal V}$ since it represents transverse momentum  (defined by Eqn(5)) transported by the $U$-stream;  It  does contribute to the local energy-momentum tensor.
  
      \subsection{ The  Streams' Action Principle.}                                             %pg 8
 
            Eqn(5) is really too restrictive to be used as a starting point, for once $\bar a,\bar b$ are given  specification of any one of the three vectors  determines the other two (in a preferred Lorentz frame).  So we delay satisfying Eqn(5).  In its place we consider two independent streams $U$, time-like, $S$, space-like, with $U\cdot S=0$ required when the two streams are very close in a small volume $\Delta {\cal V}$.
(See Appendix 4.)    The vectors are tangent to the central paths $x^\mu(\tau), x^\mu(\sigma)$, resp.  of the streams.  Again  $x^\mu(\sigma)$ 
has no physical significance  outside of    $\Delta {\cal V}$. The new twist we add is that from the point of view of a traveler along one of the  paths, say $x^\mu(\tau)$, in $\Delta {\cal V}$ when close to the other stream, $x^\mu(\sigma)$, he/she sees that stream as extended so that the other tangent vector $S^\mu$  also represents  a {\it local}  vector field parallel to $S^\mu$ representing the internal structure of the stream. (See Appendix A5.) 

           It is now fairly straightforward to set an action integral for a pair of streams and use the variation of the action to give the equations for the two different paths.    
  
   For the two stream combination the kinetic part of the action is taken to be of the form:                                                                                                               
    $$  {\cal A}_s=c^2\int \Delta x^1\int \Delta x^2  \int d\sigma  \int d\tau \ [{\cal L}_\sigma + {\cal L}_\tau + {\cal L}_{US}, ]$$
 with the variation
 $$\delta {\cal A}_s=c^2\int \Delta x^1\int \Delta x^2 \  \lbrack \int d\sigma  \int d\tau \ {\cal K}_\tau +  \int d\tau \int d\sigma \ 
 {\cal K}_\sigma \rbrack$$
where
 $$ {\cal K}_\tau= \delta[{\cal L}_\tau + {\cal L}_{US}]/\delta \tau ,\ \  \,
  {\cal K}_\sigma= \delta[{\cal L}_\sigma + {\cal L}_{US}]/\delta \sigma  $$
  
  (and  $d\tau d\sigma =-d\sigma d\tau$ if we regard differentials as one-forms).  We use
  $$ {\cal L}_\tau = m_0(x)\ [g_{\mu\nu}U^\mu U^\nu -1], \  \ \ {\cal L}_\sigma = m_1(x) \ [g_{\mu\nu}S^\mu S^\nu +1], \ \ {\rm  and}$$
 $$ {\cal L}_{US}=m_2(x) \  g_{\mu\nu}U^\mu S^\nu.                           \eqno(A5)  $$
  Here $m_0,m_1,m_2 $ are arbitrary functions, Lagrange multipliers ( of dimensions of $ML^{-4}$); $m_0$  [or $m_1$] need be  defined only along the path $x^\mu(\tau)$  [or  $x^\mu(\sigma)$]. After the variation of the action is done we shall restrict them by requiring 
  $( m_0 U^\mu)_{;\mu}=0, \ (m_1S^\mu)_{ ;\mu}=0$, the mass conservation conditions along the paths.  Also  set $m_0(x) =\bar a \ m(x)$  and $m_1(x)=\bar b \ m(x)$ to permit establishing  the momentum conservation  conditions of  Eqn(5).
  
       We set $m_2 \equiv \surd (m_0\ m_1)f(x)$, where  $f(x)$ is arbitrary. Containing a cut-off factor, it determines the size of 
  $\Delta {\cal V}$ (by the requirement $m_2=0$ outside) and, as we shall see, specifies the local field shapes carried by $S^\mu$ and $U^\mu$.

  \subsubsection{The Geodesic Equations.}                                          %pg 9                                                                                 
 
    The geodesic equations  specify how the two streams entangle within $\Delta {\cal V}$. One has, for the specified Lagrangian:
$$  m_0(x)d {\hat U}^\mu /d \tau= {\hat U}^b {\hat U}^\mu_{; b} =g^{\mu  a}(m_2  S)_{[b,a]} {\hat U}^b ;        \eqno(6a)$$
$$  m_1(x)d{\hat S}^\mu/d \sigma= {\hat S}^b {\hat S}^\mu_{; b} =g^{\mu a}(m_2 U)_{[b,a]} {\hat S}^b .          \eqno(6b)$$
                                                                                                                                                          
Here, we have used the notation: ${\hat U}^\mu =m_0(x) U^\mu;\  {\hat S}^\mu =m_1(x)S^\mu; \  S_{[b,a]}\equiv {1 \over 2}[ S_{b,a} 
-S_{a,b}$]; where $m_0, m_1, m_2$ are arbitrary functions.\footnote{The quantities $(m_2 S)_{[b,a]} ,\  (m_2U)_{[b,a]}$ represent differential operators on the local fields  $S_\mu, \  U_\mu$, really defined by {\it images}, since   Eqns (6a,b)  each refer to points on two different curves; for a solution  see  Appendix A5.} 
    In Eqn(6a,b) we put $m_2=0$ outside of $\Delta {\cal V}$ because the streams are entangled only inside the region.  Since orthogonality can only be defined a  small region, we may also regard disregard the solution ${\hat S}^\mu=0$ outside, when $m_2=0$ in Eqn(6b). [   $ {\hat S}^\mu$ itself  has no physical significance outside, since  $ {\hat S}^\mu$ represents the transverse momentum carried by ${\hat U}^b $. ]

        In much more conventional notation, both equations are of the form:                 
                                                            
$$ d{\vec v}/dt = {\vec v} \times {\vec B}     \eqno(7)$$                                                         

 which represents the motion of an ion  in a magnetic field.  Choosing an orientation such that ${\vec B}$ only has a $z-$component, $B$, the usual `guiding center' approximation is to take the helical motion as disjoint; {\it e.g.} $v_z={\rm const}., v_x=v_0 \sin B t, v_y=v_0 \cos B t$ , treating $B$ as  not varying much in a gyro-radius.  We use this `guiding center' approximation taking the spin rate as fast. We also take out a common factor $m(z-U^z\tau)^2$ [see Eqn(5)]  from  the two equations and look at the remaining axis-symmetric representation of the helical motions.

    This guiding center approximation strictly requires that $({\hat U}^0)^2 -({\hat U}^z)^2,\ ({\hat S}^z)^2 -({\hat S}^0)^2$ are constants;  these vector components will be taken as slowly varying functions of $t,z$ so that the spinning is confined to the $x-y$ plane.  [In a more general Lorentz frame $S^0$ need not vanish.]  A simple solution is discussed in  Appendix B2.  A summary is that the two streams, $U,S$  need not intersect but` braid' similar to the helices in  DNA, the factor $m(z-U^z\tau)$ determining the `length' of the braid.  However, the radii of gyration of the two streams are quite different: $\varpi_S =\surd k \ \varpi_U$.  Each point on the $x^\mu(\tau)$ helix can be associated with its `opposite' point  on the $x^\mu(\sigma)$ helix. The function  $f(x)$ in $m_2$  determines the effective value of   $B$ within  $\Delta {\cal V}$,   the  narrow sheath  around the braided streams; it is related to the streams' internal structure.   
      
\section{The  Orbital Motions Of DM For a Schwarzschild Metric.}           %pg 10                       
      The DM discussion so far is for the momentum distribution of  `braided' streams  in a local neighborhood.   Instead of solving Eqns(6a,b) directly in the vicinity of a central attractive source we adapt  the  local guiding center approximation,using the `$z$'  direction to correspond to the mean path a braided pair follows in  the gravitational field.     
         We assume the Milky Way galaxy is  represented by an exterior Schwarzschild  metric and its halo by a DM interior Schwarzschild metric. One has: 
     $$ d\tau^2 =B(r)dt^2 -A(r)dr^2 -r^2 d\theta^2 -r^2 \sin \theta^2d\phi^2,    \eqno(10)$$
    where $ A^{-1}\equiv 1-2G{\cal M}(r)/r \equiv 1-2\Phi$, and  we write     $B\equiv 1-2\Psi$, its precise form being discussed in Appendix C. For the  exterior Schwarzschild  solution, one has $AB=1$  and  $\Psi=\Phi$.
   
\subsubsection{ The Integrals Of Motion. }       
        First we show  DM  cannot approach the inner regions of a gravitational field  represented by a Schwarzschild  model.

             Use the guiding-center approximation  for representing the local motion of a braided stream pair.  If we average over the spinning,  the mean value of $U^\mu$ defines the  both the axis and the geodesic along which they translate. Call the tangent to this geodesic $V^\mu$ and take this local axial direction   to lie  along the geodesic curve $dV^\mu/d\tau =0$. At a point $ (r,\phi, \theta)$, suppose the tangent vector components are $(V^r,V^\phi,V^\theta)$.  We take $V^\theta =\langle U^\theta \rangle=0 $ to simplify the algebra. Then there is a plane of orthogonal vectors $W$ given by  $w( {\hat w}_r, {\hat w}_\phi, {\hat w}_\theta)$  where $W^\mu W_\mu =w^2$ and
      $$ {\hat w}_r= -(v_t/v)\cos \alpha, \ \ ( r \sin\theta) \ {\hat w}_\phi = (V^r/v) \cos \alpha  , \ \ r \ {\hat w}_\theta =\sin \alpha \eqno(11a)$$
      with 
      $$v_t =V^\phi r \sin \theta,  \ \  v^2 = v_t^2 + (V^r) ^ 2;   \eqno(11b)$$                           
      here, $w$ and $\alpha $ are arbitrary.   We may use this notation to represent the spinning portion of $U$  by setting 
      $\alpha =\varphi(\tau)$  and $w \to w_u =\varpi_0 U^\varphi$ (in  cylindrical notation)\footnote{The values of $w_u, w_s$ are also given by the expressions for $U^y,S^y$ in  Appendix B2.}  [And we may treat $S^\mu$ similarly using  $w\to w_s = (b/a)w_u$ and $\alpha_s =\alpha_u + \pi.$ ]  Then one has $d V^\mu/d\tau= d \langle (U^\mu-W^\mu)\rangle/d\tau =0$ as representing the usual  restriction for a tangent  vector along a geodesic. The  four geodesic  integrals of  motion for $V$
    $$ L_v= r^2 \sin^2 \theta V^\phi; \ \ J^2_v =r^4 (V^\theta)^2 + L_v^2/\sin^2 \theta;$$
    $$ \epsilon_v =V^tB; \  \ \\-N_v =A(V^r)^2 -\epsilon_v^2/B +J_v^2/r^2; $$
    written in terms  of the components  of $U, W$   become:    
       $$\epsilon_v \equiv U^t B; \eqno(12a)$$
       $$ L_v\equiv  r^2\sin^2 \theta \langle (U^\phi -w_u  {\hat w}_\phi) \rangle \  {\rm or}
\               \bar L_v  \cong   r^2\sin^2\theta \ \langle U^\phi \rangle ;  \eqno(12b)$$          %pg 11
   $$J_v^2 \equiv r^4[\langle(\delta U^\theta)^2 \rangle +w_u^2\langle {\hat w}_\theta^2\rangle]  + [ \bar L_v^2 +r^4w_u^2\langle {\hat w}_\phi^2\rangle ]/\sin^2 \theta ; \eqno(12c)$$
                  $$-N_v \equiv A[\langle U^r\rangle^2 +w_u^2\langle {\hat w}_r^2\rangle] -\epsilon_v^2/B +J_v^2/r^2. \eqno(12d) $$
       
       Here $\epsilon_v, L_v,J_v,N_v$ are constants of integration,and we  put   $\langle \cos \alpha\rangle \cong 0$ and  $\langle \cos^2 \alpha\rangle \cong 1/2$. Normally one orients the coordinate axes such that  $\sin \theta =1$ with $\langle U^\theta\rangle=0$. We assume $\langle(\delta U^\theta)^2 \rangle=0$.  Then Eqn(12c) requires $w^2_u$ to be of the form: $w^2_u=2W^2_u/r^2$ where $W^2_u$ is a constant. Putting  $N_v =1$ and introducing the constant orbital energy ${\cal E}$ by $\epsilon_v^2 =1 +2{\cal E}$, one  rewrites (Eq12d) in the more familiar form:
                                                                                                                                                               
       $$ (U^r)^2 + W_u^2/r^2 +  L_v^2/r^2 -2\Psi \cong 2{\cal E}. \eqno(13) $$
        The orbital motion described by Eqn (12d) is controlled by the form of $B$ adopted and not by the form of $A \approx 1$  (to an accuracy of $1$  part in 
        $\sim 10^{-6}$  when we are far from a central singularity).  The reason we get the extra angular momentum term in Eqn(13)
        is because  $U^\mu$  does not follow the classical geodesic; we do not have  $dU^\mu/d\tau = 0$ for all components of the motion.  
        
        There is an equivalent discussion for the components of $S^\mu$.  One introduces a vector 
        $\bar V^\mu= \langle S^\mu -W_s^\mu \rangle$, etc. Both $V^\mu$ and $\bar V^\mu$  may contribute to the energy momentum tensor.

             While both bound and unbound paths are permitted by Eqn(13) we shall consider only  unbound motions, $2{\cal E} =V_0^2 \geq 0$.
             [ Bound paths would require an additional mechanism for specifying DM energy loss.]
        The value of $V_0^2$ is determined at very large distances $r =r_q$ where  $\Psi \ll V_0^2.$ Then normally one puts $(U^r)^2 =V_0^2 = \langle  v^2\rangle/3$ the mean squared speed representing  turbulent motions.  We expect   the mean transverse `spin' velocity $W^2_u/r_q^2=w^2_u $ to be also set by the local turbulent velocity, $w^2_u  \leq(2/3) \langle v^2\rangle$, because the spin  term is always dynamically coupled to the motion.  In effect we shall assume that the spin energy of DM in a thermodynamic enclosure equilibrates with the translational degrees of freedom.
        
         Both terms contribute to determining the minimum value of $r=r_p$.   For example, for the Milky Way galaxy (with $\surd \Psi  \simeq 200 \  {\rm km/s}\ {\rm at} \ \sim 8\ {\rm kpc}$)  an entering OM object with $v_{\rm radial}\cong 0, v_{\rm transverse} =20\ {\rm km/s}$ at $R=50$ kpc will pass within $\sim 3.3$ kpc of the galactic center, but a DM object which  has additionally an internal `spin' velocity of $w_u =25 \ $ km/s at $R=50$ kpc can only approach within $10$ kpc of the galactic center.    
    
       In order for such objects to usefully represent the DM halo of the MW galaxy, which starts at $\sim 4-8\ $kpc, one requires the halo DM  to possess  very low spacial velocities as well as very low spin velocities, $w_u \leq \sim 33\ $km/s at large distances,  $R\geq50 \ $kps.  This is `cold' DM.  For the DM  halo to  have a fairly sharp inner boundary  the spin velocity should be dominant. If these velocity limits are exceeded, the galactic halo forms further out.
       
        In summary, unlike ordinary matter, DM carries along with itself an `spin'  energy term, ${\cal W}={1\over 2}W_u^2/r^2$. This contributes to the centrifugal barrier the DM experiences and prevents its orbit from approaching as close to the central attractor as the orbit of  OM does.   All DM orbits are taken to have positive energy; the braided streams cannot be captured.  One may consider that   the effective potential  is $\Psi_{eff} =\Psi -{\cal W}$  with the orbital speed (corresponding to  $u_z$ of Appendix  B1)  being given by   $v^2=(U^r)^2 +  J_u^2/r^2$.

      [ The integrals of motion for $S^\mu$ are of precisely the same form (with $N=-1$).  Since the Lorentz frame is that of the central source being at rest, one must allow for a $S^0$ term; we shall assume it to be small. Also because $w_s <w_u$ ,one has $W_s <W_u$ , and at the turning point   $U^r(r_p) =0$ one has, in general $(S^r)^2(r_p) >0$.  In our model the path $x^\mu(\sigma)$ has only significance in the vicinity of $x^\mu(\tau)$ and its extension to values of $r<r_p$ should be ignored.  The portion of $S^\mu$, representing motion perpendicular to the spin (and called  $S^z$ in Sec 3.2)  is given by $(S^z)^2= (S^r)^2+J_s^2/r^2$. [In the usual cases of interest since $W_s$ is small and $(S^z)^2$ is large compared to  $\Psi$, the approximation  $(S^z)^2   \cong$ constant is good at  distances from the central attractor much larger than the Schwarzschild radius.] 
        
 \section{Description Of The Halo Model.}      %pg13

      We regard a galactic halo as the response of the local intergalactic medium to the presence of a concentrated  OM gravitational source, similar to the Debye sphere surrounding an ion in a plasma.   Introducing  a spherical coordinate system, the  background model used  for $r>r_q$ is an  Einstein cosmology {\it i.e.} using a Robertson-Walker metric with a non-zero cosmological constant. This represents the local intergalactic medium.  For $r < \ \sim r_p$ we assume a central singularity, characterized by the exterior Schwarzschild metric; this represents a galaxy. For the galactic halo, $ r_p \leq r \leq r_q$, we use an interior Schwarzschild metric ,   with an energy momentum tensor specified  by DM only.  The first task is to calculate this  DM  energy-momentum tensor. The second is to calculate from this the density and pressure in the halo (see Appendix B3). The actual halo  structure, the values of the metric components, is then  outlined in Appendix C. 
          
        \subsection{The    DM Energy-Momentum Tensor For a Braided Pair. }   %pg 14                                                                                               
        
     As before in dealing with ordinary matter, we  consider only streams that do not terminate in the volume under study, so $ {\hat U}^\nu_{;\nu}= 
     {\hat S}^\nu_{;\nu}=0$, expressing mass conservation.  For the solution of Eqns(6a,b)  one finds that the two equations effectively differed  only by  multiplicative constants (the square of the scale lengths).  Since $T^{\mu\nu}$ is found by summing all streams within a small volume, the distinction between the two paths is not relevant. Correcting for this scale difference, the difference between the two  equations vanishes. [See Eqns(B5a,b).]  Therefore, we consider as (a  fundamental part of) the energy momentum tensor for  a single braided stream:
     
    $$ \epsilon (a^2+b^2)a^2m_0{\tilde T}_s^{\mu\nu} = a^2 \ {\hat U}_s^\mu{\hat U}_s^\nu -b^2 \ {\hat S}_s^\mu{\hat S}_s^\nu\equiv a^2m_0 \ [m_0U^\mu_s U^\nu_s-m_1S^\mu_s S^\nu_s],           \eqno(14a)  $$    
                 
           where the multiplicative constants $a,b$ satisfy $a^2m_0\equiv b^2m_1$.  [ For halo DM , $m_0=k\Gamma m,\ m_1=\Gamma m$ and $a^2 = \Gamma, \ b^2= k \Gamma$. See Sec 3.1.]  Also, $\epsilon =\pm1$;  provisionally,we choose $\epsilon =+1$, so that $\tilde T \equiv m >0$.            
           Using the coordinate system used in discussing the guiding center solution, one finds $
          { \tilde T}^{xx}={\tilde T}^{yy} =0$ so that the `spinning' motion in the $xy-$plane does not act as a source term  for a gravitational field.   [This reflects the fact that  $U_s,S_s$ were designed to be a representation of the vector $K_s$, Eqn (5), which has no $x,y,$ components in this frame.] Also, $\tilde T^{\mu\nu}_{s\ ;\nu}=0$ follows because of the requirement   $ {\hat U}^\nu_{s\ ;\nu}={\hat S}^\nu_{s\ ;\nu}=0$. 
       
      [ For OM carrying small amounts of transverse momentum, one has $m_1/m_0 = v \ll 1$, (see  Appendix A6), and we can exclude the $S^\mu S^\nu$ terms, recovering the form used in Sec. 2.  For DM  one has $m_0/m_1 =k \ll 1$  and to good approximation we can exclude the $U^\mu U^\nu$ term.]       
    \subsubsubsection{ An Averaged Local Energy Momentum Tensor.} 
       
      In a local Lorentz frame the only stream components contributing to ${\tilde T}^{\mu\nu}$ are: $U_s^0,U_s^z,S_s^0,S_s^z$.     For `adding' isotropic ensembles  of similar DM streams  in a small region  to form  volume {\it field} averages  we can follow the same procedures used  in Sec. 2. From the variance in the distributions of ${U^\mu_s}$, we get conventional  pressure and internal energy terms expressed by  the tensor $P^{\mu\nu}$  of Eqn(3b).  For simplicity we shall  ignore these terms.  One finds the leading terms\footnote{We can set a particular $S^0_s \equiv 0$ only in one rest frame.  So for an ensemble of braided streams one really get an extra term $\delta \tilde T^{00} =-\bar m_1 {\cal N}\langle (S^0)^2\rangle $ which we will assume is small.}  are:
           
      $$\langle \tilde T^{00}\rangle   =  + \bar m_0{\cal N}\langle a^2(U^0)^2 \rangle, \ \ \langle \tilde T^{xx}\rangle =   \langle \tilde T^{yy}\rangle =\langle \tilde T^{zz}\rangle \cong -\bar m_1{\cal N}\langle b^2 (S^z)^2 \rangle/3 ,   \eqno(14b) $$
                                                                                                                                                                 %pg15
  with $\langle \tilde T^{\mu\nu}\rangle =0$ otherwise. Here $ {\bar m}_0,\bar m_1$  are average stream mass densities; ${\cal N} $ is the average number of braided streams in a unit volume; and $ \langle(S^z)^2\rangle$ is an average of the squared value of  that component of $S$ which is normal to the spin-plane in the guiding center solution.  
  
  Neglecting contributions from the variances of the motion distributions, we do not have a real pressure term.   The ensemble average may be represented by the usual  ideal fluid representation $ T^{\mu\nu}_{fluid}  = (\bar \rho + \bar p) \bar U^\mu \bar U^\nu -\bar p g^{\mu\nu}$, but now $\bar p$ is {\it negative}.  The DM fluid acts to assist compression and resist expansion.   The term $\bar p$ is not really a pressure, but historically $ T^{\mu\nu}_{fluid} $ has been treated as if it were representing a real fluid.  We shall refer to $\bar p$ as the {\it false} pressure, 
 and  similarly $\bar \rho$ as the { \it false} density.

 \subsubsubsection{ The Alternate Local DM  Energy Momentum Tensor For Cosmology.} 
 
    We now show that a conventional  form for $T^{\mu\nu}$, with an effective positive $\bar p$ is possible if we add extra terms, including a cosmological `constant' term to the source terms.    First consider
            $$ T^{\mu \nu}_{alt} \equiv \langle \bar T^{\mu\nu}\rangle +  q _a \langle V^\mu V^\nu  \rangle      \eqno(15a) $$
     where $V$ is that component of $U$ without spin normalized so that $ V^\alpha V_\alpha=1$; it satisfies $V^{\mu\nu}_{;\nu} =0$ ( See Sec 4.0.1. ) Choose     $q_a = |\bar p |/  \langle (V^0)^2\rangle$.  
Then we get rid of the anti-expansion term  by using some subterfuge. Let's define  
   $$T^{\mu\nu}_{dark} \equiv  \ T^{\mu\nu}_{alt} -\lambda(p) g^{\mu\nu}. \eqno(16a)$$
       
     One has that $T^{00}_{dark} = \rho + q_a \langle (V^0)^2\rangle  -\lambda \equiv \hat \rho $ and $T^{zz}_{dark} \cong p +q_a \langle (V^z)^2\rangle +\lambda \equiv \hat p$.  Choose $ \lambda = - p >0 $; Then $T^{00}_{dark}=\rho= \hat \rho >0$ and {\it e.g.} $T^{zz}_{dark} =q_a \langle (V^z)^2\rangle =  \hat p >0$ .  We may now include the usual pressure and internal energy terms  associated with $P^{\mu\nu}$, expressing  the variances in the velocity distributions. (See Sec.1.) In the local rest frame all terms of $T^{\mu\nu}_{dark} $  are diagonal and positive and we can therefore write
     
  $$T^{\mu\nu}_{dark}= (\hat \rho + \hat p)U^{\mu\nu} -\hat pg^{\mu \nu}  \eqno(16b)$$
  
as for a normal fluid.   But then we must rewrite  Einstein's equation as
$$  R^{\mu\nu} -g^{\mu\nu}/2 =8\pi G \langle T_{alt}^{\mu\nu} \rangle=8\pi G\  T^{\mu\nu}_{dark} +\Lambda g^{\mu\nu},  \eqno(15b)$$

where we have put  $\lambda =\Lambda/(8\pi G) $, regarding this term as part of the rest of Einstein field equations  not obviously associated with the source term.  It is necessary to  add the extra term  because  $\lambda$ is not a constant but  really varies as $p$ does. It's variability makes no difference because in Eqn(15b) we have added and subtracted the same term in the RHS of the equation.  So we have a `normal' fluid representation  of the dark matter providing we also introduce a cosmological term into Einstein's field equations.   We suggest that this is the form implicitly used for representing DM in constructing our  standard  cosmological models.     
\subsubsubsection{ A Criterion For $T_s^{\mu\nu}$,  Allowing  The Choice $\epsilon < 0$ for DM. }   %pg16

        Since symmetric tensors can be added to a proposed  ${\tilde T}_s^{\mu\nu}$ a criterion for an acceptable form of it would be useful; $\tilde T^{\mu\nu}_{s;\nu}=0$  is not sufficient.
        Starting with an initial stream momentum vector $\hat K_s^\mu$, consider a local isotropic ensemble of similar streams,  for which the orientation of the space part of the vector is random, and form the average of the proposed  ${\tilde T}_s^{\mu\nu}$ form. [ Equivalently, consider a representation of  $\hat K_s^\mu$  averaged over all possible spacial orientations of the local Lorentz frame.] Then $\langle \tilde T^{\mu\nu}\rangle $ is diagonal  with equal space components.  Add $\lambda g^{\mu\nu} $ to it where $\lambda = \langle \tilde T^{zz}\rangle$. 
        The resulting sum $T_E^{\mu\nu}$ then has only one component, $T_E^{00}$ ;  in the case of the stream source term of Eqn(14a), one finds
        $$ T_E^{00} =\epsilon {\cal N} (m_0-m_1/3) \equiv g_{\mu\nu}T_E^{\mu\nu}.    \eqno(17)$$
        
        This construct emphasizes Einstein's  original focus on the $T^{00} =\rho_{eff}$ as the  source of the gravitational field with the other components of $T^{\mu\nu}$ ignorable. [Einstein {\it ``The Meaning Of Relativity}, 5th ed.] For an attractive gravitational field, using Eqn(14a) one must choose $\epsilon =-1$ for DM if $m_1 > 3m_0$.( Or, one must add an additional term to Eqn (14a).)  The ideal fluid representation for an ensemble of DM streams  represents a source of possible confusion; $\bar \rho,\bar p$ need not be functionally related, but simply constrained by $\tilde T^{\mu\nu}_{s;\nu}=0$.  We cannot use this construct  directly in all applications because $\lambda \neq $ constant.

        We adopt Eqn(14a), with $\epsilon =-1$  for representing DM in the remainder of this paper. We have $ T^{\mu\nu}_{fluid}  = (\bar \rho + \bar p) \bar U^\mu \bar U^\nu -\bar p g^{\mu\nu}$, but now  the false pressure $\bar p $ is positive and the false density $\bar \rho$  is negative. We have neglected contributions from the variances of the various velocity distributions.

  \subsection{ Representation Of The Averaged $T^{\mu\nu}$ In The DM Halo Model.}          %pg17
       In this  very simple halo model we assume no OM and only one type of DM, that for which the energetics at $r \geq r_q$ are given by $K: \Gamma(k,0,0,1)$ in a local Cartesian frame ; See Eqns(5).  The DM  follows orbits satisfying the integrals of motion; see Eqns(9a,b,c,d).  We choose  $r =r_p$ to be the turning point of  a representative stream  $U^r (r_p)=0;$  at that radius,  $(S^r(r_p))^2 >  0$ (See  Section 4.0.1).  For $r \leq r_p$ the solution for $S^\mu$ has no physical significance and one may take $T^{\mu \nu}=0$.  
       
         In the annular region $r_p \le r  \leq r_q$, one specifies the halo $T^{\mu\nu}$ by the change of the halo $T^{\mu\nu}$ from its value in the background model at  $r= r_q$ caused by the compression of the pressure in this inner region. For DM one has $p > |\rho|$.        
      We may write  for the typical  spacial diagonal component  of $\delta T^{\mu\nu}$ at the  halo point  $r_a\geq r_p$,  
      
      $$\delta T^{xx}_a = {\bar m}_a\  {\cal N}_ a {\bar P}_a - {\bar m}_q\  {\cal N}_ q {\bar P}_q.   \eqno(17)$$
      
         Here ${\bar m}$ is an average stream density, ${\cal N}$ is the number of streams in a unit volume and ${\bar P}$ represents an average value of the square of the `speed'  $s^2\equiv (S^r)^2   +  J_s^2/r^2 $ along the paths $x^\mu_s (\sigma_s)$;  it corresponds to$ \langle(S^z)^2\rangle/3$  of Eqn(15). Using the steady state one dimensional pipe flow; one has $ {\bar m}_q / {\bar m}_a = s_a/s_q$.   A similar expression holds for $\delta T^{00}$ in terms of the average value of $(U^0)^2$. 
       
        With these considerations,  the radial dependence of  $\delta p(r), \delta \rho(r)$   can be found to represent a spherical halo resulting from infall from intergalactic space.  See Appendix B3.  Using these values for the source terms,  the radial variations of the metric elements $A(r), B(r)$ are given in Appendix C. The predicted halo rotation curve is given by Eqn.(C3).
 
 \section{Results.}      %pg 17
 
      A `flat' rotation curve {\it is}  the Einstein equation, Eqn(C3), for one of the metric coefficients when the pressure term $p$  (1) is $\propto 1/r^2$; and (2) dominates over the classic Newtonian potential $\Phi(r)$. Its applicability is confined to the outside of galaxies by an assumed property of  DM. We require that its  constituents  each have intrinsic angular velocity which, combined with  orbital angular velocity,  form     a centrifugal barrier to prevent close approach to the center of attraction.
      
   All galaxies should have halos. A model halo starts at $R_h=r_a^2v_a^2/G{\cal M}_0$; an upper limit to the extent of these models is set by the gravitational fields of its nearest neighbors.  A low mass galaxy should have its DM halo starting well beyond its visible structure. Ordinary matter entrained with the DM can fall into the region $r<R_h$.  A high mass galaxy may have its structure begin inside its  visible structure. The Milky Way  probably is such a galaxy.  So far, observations of other large spirals suggest  values $R_h \sim 4-8$.                                 
      
      If the DM  has low orbital angular momentum, there is an outer  halo `free-fall' zone and an inner halo zone in which conservation of angular momentum controls its radial velocity.  For DM this inner zone has a lower bound specified by the intrinsic angular momentum of the DM; it cannot enter the  region in which the main bulk of the galaxy's OM resides.  While one expects OM to constitute $\sim \Omega_b/\Omega_d \sim 1/6$ [16] of the halo density at large $r$ it  does not have intrinsic angular momentum and will fall through the halo boundary.

        In the halo's large outer free-fall zone, $r_b^\star <r < r_a$, where  the `false' pressure, $p \propto 1/r^2$,  is dominant, one finds that   the observed $v_{cir}$ is  nearly constant .  See Eqn (C3b).  The exact departure from strict constancy depends upon the contribution from the potential of the central galaxy mitigated by (negative) contributions  from DM density terms.         In the inner halo zone the rotation curve is given by Eqn(C3a) and $p \cong$ constant. In this region the transition from the central galaxy's rotation to the halo's rotation may be quite abrupt, because the zone may be relatively small in extent.  %pg 18

        As an example, we construct a simple representation for the Milky Way galaxy's rotation  curve for $r \geq 6$ kpc.   For the galaxy, we assume  $V(r)^2 =G{\cal M}_0/r$, with $V(6 \ {\rm kpc})   =220\ {\rm km\ s}^{-1}$.  The halo model  uses $R_h=8$ kpc as the effective edge of the halo,   with  $\eta =1/2$ (so no DM can go below $\sim 4 $ kpc and the boundary between the inner and outer halo zones is at $r_b^\star =12$ kpc). We use the  lowest order approximations\footnote {These  greatly simplify  the the detailed halo calculations in the range $r= 4-7$ kpc  where the halo contributions are very small and are good approximations in the range $r=9-12$ kpc. }   discussed in Appendix B3; in this case one sets $v_h^2 =4\pi G(p_a+\rho_a)r_a^2$ for $r >R_h$.  We chose $v_h  = 175\ {\rm km\ s}^{-1}$. Then for $r=8,12,20,48$ kpc one gets, 
 using $v_{cir}^2 =V(r)^2 +v_h^2 $, that $v_{cir} \cong 190, 230,210,190\ {\rm km\ s}^{-1}$, resp., with $v_{cir} \to v_h$ at larger $r$. This represents the `flat' portion of the rotation curve, the observed flatness  resulting from using a straight line average of the observations. ({\it e.g.} See [17].)

     This agrees with the  MW rotation curve points  depicted in [18].   There is some leeway.  For the Milky Way galaxy, a spatially averaged rotation curve is not available and it is known that some inner regions on opposite sides of the galaxy exhibit differing rotation curves [ 15].  Also, estimates of the luminous mass ${\cal M}_0(r)$ differ by a factor of two [18,19,20,21 ].
A fit  to the data used in the rotation curve of [17]   would favor  $R_h=6$ kpc. [ The above calculation used $k \equiv  | \rho |/p=0$. Use of $k$ up to 0.15 gives about the same results. ]                                                                                                                      %pg18

  Using $v_a \sim 30 \ {\rm km \ s}^{-1}$, the rotation curve parameters used correspond to the theoretical value  $p_a/c^2 \sim 1.5 \times 10^{-26} \ {\rm g \ cm}^{-3}$ at $r_a \sim 50 $ kpc or $p_a/c^2 \sim 0.5 \times 10^{-26} \ {\rm g \ cm}^{-3}$ at $r_a \sim 80 $ kpc. We have no good way of choosing either value $r_a$, but both estimate for $p_a/c^2$ looks reasonable. [ The smaller distance corresponds to an assumed shock radius at which we have suggested $p_a \sim 1-10\  p_q$ where $p_q$ is the effective density of DM in intergalactic space.] These should be compared to other appropriate  densities [23,24]. The limiting cosmological critical density is $\rho_c \sim 10^{-29} {\rm g \ cm}^{-3}$. The mean densities of luminous matter in clusters of galaxies are $\sim  10^{-26}\ -\ 10^{-28} \ {\rm g \ cm}^{-3}$;  This is on the order of our estimate for $p_q$.  
  
     The false pressure in the inner halo $p_b=p_a(r_a/R_0)^2 \approx 50-100\ p_a$ corresponds to a density $\sim 10^{-24}-10^{-25}\  {\rm g \ cm}^{-3}$ which is comparable the OM  mean density of the MW galaxy distributed inside a sphere of  $10$ kpc radius,   $\langle \rho \rangle \sim 10^{-24} \ {\rm g \ cm}^{-3}$.  For $p_b > \sim \langle \rho\rangle /10$ we found that the rotational velocity does not decrease in the inner halo region accounting for the observed  abrupt transition in the galaxies' rotation curves.

\section{Discussion.}                                                             %pg 19
\subsubsection{ The Need For Observations.}   

Of the three needed quantities, $p_a,r_a, v_a$, (with  $p_ar_a^2=p_b(r_b^\star)^2$) present observations of spirals give $R_h=r_a^2v_a^2/G{\cal M}_0$ and $v_{cir}(r_a)=\surd (4\pi G p_ar_a^2)$; they do not determine $r_a$ or $p_a$ separately. Here, $v_a$  is the mean speed of the DM peculiar velocities 
at $r=r_a$. Then $v_{cir}^2/R_h $ determines ${\cal M}_0$ or $p_a/v_a^2$  if the other is known. Often ${\cal M}_0$ can be estimated from the rising portion of the rotation curve.aries on a cosmological time scale.
                                  
       There  are two choices for $r_a$. The larger is something smaller than  half the mean distance  to nearby substantive galaxies where the potential of the central galaxy dominates over its neighbors.   Observations of DM lensing [22 ] in earlier epochs are modeled with $r_a >100$ kpc. In the case of the MW galaxy,, which exhibits a  flat rotation curve at $\sim 200\ {\rm km \ s}^{-1}$, one has limits set by  disturbances at $\sim 60$ kpc   by the  dwarf galaxies LMC \& SMC, and at $\sim 700$ kpc by M33; evidently the assumption of a spherical gravitation potential would would be crude and useful only in selected directions. A study of MW `halo' objects suggests the halo changes character at $\sim 30$ kpc and might extend to $\sim 120 $ kpc [13]. 
        
        In general there is a problem in using for $r_a$ a  large fraction of the mean spacing ($\sim 10^3$ kpc)  between galaxies since one must have  $p_a > \rho_c$, the critical cosmological density.  The value of $p_a$ actually  should be set by local cosmological evolution.  Choosing a very large $r_a$  would cause the observed `flat' portion of the rotation  curve $v_{cir} \propto r_a \surd p_a$  to be very large.  Consequently, in text we have favored the view that $r_a$  is set by the stagnation radius of a discontinuity  traveling in front of a moving galaxy and used  $r_a \sim 5-20 \ R_h$ when $v_{cir} \geq \sim 200 \ {\rm km \ s}^{-1}$.     
        
        This shock proposal needs verification. What are urgently needed are observations of  galactic halos at  very low surface brightness.  Because OM and DM matter are mixed, the shock discontinuities occurring because of the galaxies' motions through the intergalactic medium should be illuminated by $H\alpha$ emission. Such observations could determine  $r_a$, and possibly $p_a$ and $v_a^2$.
        
          The deduction $p_ar_a^2 =p_b(r_b^\star)^2$ can be indirectly tested.  [ It holds only if the DM is truly non-reactive and the attracting galaxy has a spherical symmetric potential.]  The future reduction of the GAIA observations should be able to establish $V(r)$ and the  Oort constants $A,B,C,K$ ,amended by $\partial V^i/\partial z$, in the annulus $r =8 \pm 2$ kpc both in the plane and at heights $\sim 1$ kpc above it. This should allow separation of the galactic and the halo potential fields  and  specification of the DM interior halo structure with a determination of the parameter $\rho_b(r_b^\star)^2$; this would establish a value for $k\equiv \rho_b/p_b$. We have assumed   $k $ is small in calculating  rotation curves and that it is an intrinsic property of undisturbed DM.    Also more accurate observations  of other galaxies' rotation  curves corresponding to  the inner part of their halos could be used to calibrate the value  of the DM density.

   We see no reason to assume $p_a,v_a^2$ are nearly constant or obey scaling laws reflecting spherical symmetry in the galactic halos exhibiting gravitational lensing because there is no obvious mechanism to enforce these requirements in the outskirts of very large systems in reasonable timescales.  One also notes that because of past galaxy-galaxy collisions, it would be impossible to rule out the presence of DM inside  galactic disks and their possible contribution to the broad disk population and to spiral arm densities.
   
 \subsubsection{ Theoretical Considerations.}     %pg 19/20
 
       This model of the DM source term requires two  parameters: (1) $\eta$ which specifies a relation between the internal rotation velocity and the r.m.s. thermal velocity; and (2), $k= |\rho |/p$. It is possible that $k$ varies on cosmological time scales. 
       If the suggested cosmological representation  $T^{\mu\nu}_{dark}$ is valid,  then  $\Lambda \sim p$ is not constant and the Standard Model      needs re-examination.  The observed [16,24 ] cosmological parameters of  the Standard Model  $\Omega_d/\Omega_\Lambda \approx 0.15/h^2 \sim 0.3$   would then represent an averaged $k$  ; it would very likely represent the value of $k$ when nucleosynthesis was frozen out, $T\sim 1-5$ Mev. Our (preliminary) calculations for the MW rotation curve suggest $k \leq 0.15$ presently.
       
     The unusual stream approach adopted in Part 1 was adopted because it directly gives a formulation for $T^{\mu\nu}.$   The essence of the argument is that we can construct $T^{\mu\nu}$  by combining the equations of motion and conservation of mass for  single streams. The forms for $T^{\mu\nu}$  found for DM follow this procedure.

       A third quantity $f$, a classical `form factor',  specifies the internal structure of the streams; it gives the density of the streamlines within a stream.   If $S^\mu$ is replaced by the more conventional notation  $A^\mu$ and  $f\to e$, a constant charge, one gets the standard representation [26] of a particle moving in a magnetic field.  Indeed the requirement $S^\mu_{\ ;\mu} =0$ (or $U^\mu_{\ ;\mu}=0$), specifying that a path is neither created nor destroyed in a local region is the usual radiation gauge condition in disguise ($\partial_t \phi + \nabla \cdot \vec A =0$). What is surprising is not that $S$ acts as a vector potential for $U$ but that $U$ acts as a vector potential for $S$; this is not standard electromagnetic theory in disguise.

     From the Lagrangian formalism, a particle physicist would normally regard the term ${\cal L}_{US}$ as describing a short range interaction  with $f(x)$ determining the local `force' between  two fields. This interpretation is not strictly germane  to this model, but $f$ does determine the `spin' rate of the two braided streams. Since  $2\oint d\varphi\int (fU)_{[x,y]}\varpi d\varpi \equiv 2\pi \ \varpi^2 f{\cal F}  \neq 0$, by Stokes' theorem a  toroidal $B-$like  {\it field},  the quantity $f{\cal F}$, is produced in a sheath around the streams.  We would expect that if a third stream enters  $\Delta {\cal V}$, it would interact  with the braided streams through this field. [ In that case we note that the  helicity of the  braided stream's spin would play the role of representing the orientation of their effective dipole.]

 \section {Three Conjectures.} 
 \subsubsubsection {Should Ordinary Matter Be Represented By Braided Streams?} 
   We have not investigated  whether  it is useful to  consider the case when the vector $K$ of Eqn(5)  is time-like. The decomposition into 
     $U$ and $S$ vectors, discussed in Appendix B1 indicates that the contribution from the vector $m_1S$ would be very small. It is possible  that it might be useful to quantize $m_1S$ and try to represent ordinary particles with  non-zero spin this way.
\subsubsubsection{Is A String Theory Approach  Useful ?}   
        In this paper we considered the action integral for a single braided pair of streams. 
         If the guiding center approximation is vigorously adopted, the local variation of $U^x,U^y$ and $S^x,S^y$ is disjointed from the development of the other vector components and determined by the coordinates $(\sigma,\tau)$. It should be possible to divide the action integral into two parts, one of which represents the spin part. One would expect then  that this could be regarded as a bosonic `2-brane' in string theory and indeed if $f=0$ the action adopted is the ``Polyakov" action.         
         This would imply a rich spectrum of excitations  for the internal spin.  If one adopts the two transformations into the scaled coordinate $\xi,\eta$ discussed in  Section B2, with $s^\mu \to -u^\mu$ required, then one finds  $ d^2 \xi /d\tau^2 - d^2 \xi/d\sigma^2 =0$ (and similarly for 
          $\eta$), a standard bosonic string equation. However such a representation  would need to incorporate an interpretation for  $f \neq 0$ .

  \subsubsubsection{ Can We Estimate How Much DM Is Present? (A Cosmological Proposal.)}
  
  We introduced our model for DM by supposing there had been a state of the universe in which both space-like and time-like geodesics were permissible;  then our braided streams  carrying space-like momenta
  may have had a simpler  internal structure and could then be represented as traveling on space-like geodesics.  If this were so and   the directions of travel were  taken to be isotropic, then
  the ratio of space-like to time-like streams would be in inverse proportion to the solid angle enclosed by a local light-cone to that of the excluded region,$\int^{\pi/2}_{\pi/4} \sin \theta d\theta /\int_0^{\pi/4} \sin \theta d\theta$; this gives the frequency of space-like trajectories, the predecessor of DM, to be 2.41 times the frequency of time-like OM streams. [ If we also include the presently excluded trajectories for which $U^0 $or $  K^0<0$, the ratio becomes 5.82.]   The present  [22 ] cosmological estimate  is $\Omega_d/\Omega_b \sim 4.7$ The `agreement' is close enough to suggest this supposition is not implausible.  
       This scenario, of a breakdown in the permissibility of space-like  trajectories is easily modeled.
     First, in a local  Minkowski neighborhood consider a plane enclosing the $t-$axis and a vector $V$.
     We introduce a local coordinate system $(t,z)$ so that the vector has components $(V^0, V^z)$.  The plane intersects  the local light cone and we introduce null coordinates:
     $$ \alpha =(t+z)/\surd 2; \ \ \beta = (t-z)/\surd 2 .$$
     so $$x^\mu = t {\bf e}_t +z{\bf e}_z = (\alpha + \beta)/\surd 2 \  {\bf e}_t + (\alpha - \beta)/\surd 2 \  {\bf e}_z=\alpha  {\bf e}_ + +
     \beta {\bf e}_-$$
     One has ${\bf e}_+^2={\bf e}_-^2 =0$ and ${\bf e}_+ \cdot {\bf e}_- =1$.   The radial vector $x^\mu $ is representative of all (contravariant) vectors in this space such as $V^\mu$. So, $s^2 \equiv t^2 -z^2 =2\alpha\beta$. 
     
      We introduce a  reflection `parity'  operator:
     $$ {\cal P}
     _+ \pmatrix {\alpha \cr \beta \cr} \equiv \pmatrix{-\alpha \cr \beta \cr}$$
     which changes a space-like (time-like) vector into a time-like (space- like) vector. [ One has  ${\cal P}_+ =-\sigma_z$, the Pauli matrix.]
     
         So suppose there was an epoch  in which state vectors were not parity sensitive   ${\cal P}_+ \mid> =\mid>$  so that space-like and  time-like momenta  were equally probable for a particle state.   If  the universe changed by `abruptly' breaking this parity invariance one would expect an admixture of disjoint space-like and time-like states to result. [ It is possible this parity breaking  may be  related to the initiation of inflation.]
         
          Further, suppose  in this epoch the  the state vectors were also invariant to rotations of the null-vector basis such as
         $${\cal R}\pmatrix {\alpha \cr \beta \cr} \equiv \pmatrix{\beta \cr -\alpha \cr},$$
         [so that ${\cal R} = i\sigma_y$.]  This assumption is not trivial and introduces  another `parity'  operator. In our original space, the time reversal operator
         $${\cal Q}_t \pmatrix {t \cr z \cr} \equiv \pmatrix{-t \cr z\cr}$$
         becomes in the null-vector basis ${\cal Q}_t =-\sigma_x$;  consequently one has
         $${\cal Q}_t ={\cal P}_+{\cal R}               \eqno(5)$$
        and state vectors would be time-invariant under time reflection if parity invariance holds. 
                                                                                                                                                                        
        We have no suggestion as to what mechanism broke parity invariance  but note its timing would be certain; it would be when time began. 
   
   \section{ Summary.} %pg 23
     A  model of the DM surrounding galaxies has been developed which represents the observed rotation curves and can explain the absence of DM in  small potential wells such as the solar neighborhood.  In it the DM is treated as a new state of matter in which matter streams, confined to time-like paths, also carries considerable transverse momentum. This transverse momentum results in an intrinsic angular momentum term
 which, coupled to the usual orbital angular momentum, prevents the DM streams from passing close to centers of gravitational attraction. [See Sections 3 \& 4 and Appendix B2.]
 
     Aggregates of such DM streams are assumed to comprise the main component of the intergalactic medium.  Because of the gravitational attraction of an (ordinary matter) galaxy, the intergalactic DM medium locally compresses to form a galactic halo. Two alternate forms of the energy-momentum tensor representing these fluctuations is derived in Sec 5 \& 6.  In one, made to conform to the conventional form of  $T^{\mu\nu}$, a cosmological `constant' must be introduced.  The other features a  dominant `false' pressure term. Using the latter form, the structure equations  for the halo are developed which show that the false pressure  increasing considerably near the central attractor.
 The effect is to produce a constant circular velocity through most of the halo.[See Appendix B3.]  No ordinary matter is included in this simple modeling of the halo.  In effect, the DM halo provides the equivalent of Debye shielding of an ion in a plasma, allowing the  galaxy's gravitational field to join onto smoothly that of the intergalactic medium. [See Appendix C.]   
 
      Once the possibility of a large false pressure term is admitted, then the form of the conventional structure equation for the potential $\Psi(r)$ (see Eqn(C2a) forces a `flat' rotation curve whenever the DM is freely falling into the central galaxy. The only open question is where the free-fall begins and where  it ends.  The outer edge $r_a$ is that distance from the central attractor at which its gravitational attraction dominates over that of neighboring galaxies.  The inner edge $r_b^\star$ is when free-fall ends; it is approximately given by $r_b^\star \cong R_h =r_a(v_a^2/V_0^2)$ where  $R_h$ is the effective inner edge of the halo, $v_a$ is the mean DM r.m.s. speed at $r=r_a$ and $V_0=\surd (G{\cal M}_0/R_h)$ with ${\cal M}_0$ being the  mass of the central  attractor. %[ T
       
\section{Appendix A: Mathematical Conventions, Assumptions \& Details}         %pg 24          
             
             Generally we follow the conventions of [11]  with $g^{\mu\nu}$  having the signature (1,-1,-1,-,1) and Einstein's equation having the form
             $ R_{\mu\nu} -g_{\mu\nu}R/2= 8\pi G\  T_{\mu\nu} + \Lambda g_{\mu\nu} $. Specific expressions for the metric coefficients in various models were taken from [10].
             
                      A halo model consists of a background cosmological model and the insertion of additional small source terms in a small region
              ${\cal R}$.  Einstein's equations are assumed solved by a perturbative process, such as using the post-Newtonian procedure in the linearized equations.   Because of the perturbative nature of the solution, we also assume the boundary conditions can be successively refined.
              
 \subsubsubsection{ A1: The local Region.}              
              Take a region  of space-time for which  there is a coordinate covering  and in which the affine connections are continuous and finite. Then ${\cal R}$ is a neighborhood of an arbitrarily chosen point of this region; with a metric tensor we can assign  to ${\cal R}$  a 4-volume ${\cal V_R}$.  At each  neighborhood  point we may introduce a Local Lorentz frame where the affine connections vanish, $\Gamma^\xi_{\mu\nu} =0$.   For discussing the properties and kinematics of streams  and the local form of  $T^{\mu\nu}$ we use  for the Minkowski space $g^{00}=1,\ g^{ij} =-\delta ^{ij}, \ g^{0i}=0$.  We shall assume ${\cal R}$  is sufficiently small that  translational invariance of portions of streams (see A3 and A5) can be defined easily; we regard this as an application  of Poincar\'{ e} invariance.

\subsubsubsection{A2: Stream Characteristics}     %pg 24
                
 A stream may be regarded as a bundle of nearly coincident paths transiting ${\cal R}$. There  is a central curve $x^\mu(\tau)$ with tangent $U^\mu \equiv dx^\mu(\tau)/d\tau$ (and $dU^\mu/d\tau =0 $ if it is a geodesic). In Minkowski space for a stream one has  $x^\mu(\tau)=U^\mu_0 \tau$ where the $U^\mu_0$ are constants and $\tau$ is a path parameter. Here $x^0=ct$; normally we take $c=1$.  From the Lagrangian formalism, one sets $m(\tau)=\surd (-g) \rho$ where $\rho$ has the dimensions of an ordinary 3-dimensional mass (energy) density.  To allow re-parameterization of the path, we write $m(\tau)=m(\ x(\tau), U^\mu(x(\tau))\ ) =m(x)$. We consider only paths with $m(\tau) \neq 0$ in our region. The sign of $\tau$ is specified by requiring  $U^0>0$.  In discussing discrete streams we  assign small discrete  `cross-sections', $ \Delta^3 (x)$, to each stream; these volumes  do not `spread' in the neighborhood under consideration. [Any variation in the cross-sections along a path can be incorporated into the variation of $m(\tau)$.]  For each stream, a measure $\mu^a$ is needed to evaluate the widths, so $\int \mu^a dx^a \equiv \int \Delta x^a $ is always finite.    In other words, the vectors $ U^d $ are taken to be the co-tangents of the 3-form vector space, 
 $\Delta^3_d =\Delta x^a\wedge \Delta x^b\wedge \Delta x^c $;  this allows us to extend  the definition  of streams to include space-like streams  with well-defined  (constant) cross-sections.  In representing a stream by {\it e.g.} ${\hat U}^\mu_s =m_sU^\mu_s$, we adopt the convention that the velocity is normalized,  $g_{\mu\nu}U^\mu_sU^\nu_s =1$, with  any variations in the  actual normalization being absorbed into the factor $m_s$.  The variation of  {\it field} tensors, such  as $V^\mu$ along a stream, is given by:  $\delta V^\mu =\delta \tau \  dV^\mu/d\tau =\delta \tau\  U^\alpha V^\mu_{;\alpha}$

 \subsubsubsection{ A3: Local Addition   Of Streams By Translation.}       %pg 25
 
 In the small region ${\cal R}$, choose a point as the origin of a coordinate system and an enclosing coordinate box $\Delta^4 x$  centered on the origin. Using Cartesian coordinates, each stream (or streamline) passing through the box has a closest point to the origin, $ x^\mu_s (\tau_s^0)$, closeness  being defined by using an Euclidian metric. Then $\hat{x}^\mu_s  =x^\mu_s (\tau_s) -x^\mu_s (\tau_s^0)$ is a  curve which passes through the origin and summation (and averaging) of the translated streams' momenta can be performed there. 
 
  The variation of a {\it field}  quantity such as ${\bar U}$  is  described similarly  using   any point in ${\cal R}$ as the origin.  The sum defines tensor densities,  {\it e.g.} ${\bar U}^\mu \surd (-g)\Delta^4x$.   Then   $U^\mu_{s \ ;\mu} =0$ guarantees ${\bar U}^\mu_{;\mu} =0$.  Eqns(1a,1b) can be extended by the same technique to apply to all points in ${\cal R}$.
   \subsubsubsection{A4: Orthogonality Of  Contiguous Streams Is Defined by {\it Images}.}  %pg 25
  
          In the vicinity  $\Delta{\cal V}$  of any point, $x_0^\mu=x(\tau_0)^\mu$, of a time-like stream $U^\mu((\tau)$ there is a confluence of space-like streams ${\cal S}^\mu(x(\sigma))$.  We shall take the characteristic dimensions of   the region  to be much  larger  than the natural widths of  the streams.  Again   use local translational invariance. At $x_0^\mu$ there is a three dimensional subspace of vectors $A^\mu$ with
           $A^\mu U_\mu=0.$ [{\it e.g.} Consider the local Lorentz rest frame  with $U=(1,0,0,0)$. Then one may choose $A=(0,a,b,c)$ where $a,b,c$ are arbitrary.]   Now, consider the  `displaced' path:
  
  $${\hat x}^\mu(\sigma)=A^\mu +{\tilde x}^\mu (\sigma)$$                                              
                                                                                                                           
   where the relevant portions of $ {\hat x}^\mu, {\tilde x}^\mu$  lie within  $\Delta {\cal V}$ and $A^\mu = {\rm const.}$ ($A^\mu_{;\nu}=0$) within $\Delta {\cal V}$.  If $ {\hat x}^\mu(\sigma_0) =x_0^\mu$ and
  $ U^\mu \hat S_\mu =0$ [where  $\hat S^\mu \equiv d{\hat x}^\mu/d\sigma \ {\rm at}\ \sigma=\sigma_0$ and $\tilde S^\mu \tilde S_\mu=-1$ ], we will say ${\tilde x}^\mu(\sigma)$  is a space-like stream orthogonal to $x^\mu(\tau)$   when $\sigma = \sigma_0$.\footnote{The parameterization $\sigma$ of the path ${\tilde x}^\mu (\sigma)$ is assumed to be such that if {\it e.g.} $z(\tau) =U^z \cdot (\tau-\tau_0)$ and  ${\hat z}(\sigma) =A^z+S^z \cdot (\sigma-\sigma_0)$ describe the same local path, then $U^z=S^z$.  The interval $\sigma-\sigma_0 $ is measured in the same units  as  is the interval $\tau-\tau_0 $. If there are many such points $ {\tilde x}^\mu(\sigma_0)$ we may restrict the set by an additional restriction such as  ${\tilde x}^\mu(\sigma_0)$ is closest to $x(\tau_0)^\mu$.} The path ${\hat x}^\mu(\sigma)$ will be called an {\it image} of ${\tilde x}^\mu (\sigma)$ at $x_0^\mu$. 
    
       Also, using a local coordinate system, consider the  `scaled' path:
   $${\hat x}^\mu(\sigma)=A {\tilde x}^\nu (\sigma)$$
   where $A=$const. and  $ {\hat x}^\mu,{\tilde x}^\mu  $ in  $\Delta {\cal V}$ with ${\tilde S}^\mu{\tilde S}_\mu =-1$. If one has  that 
   $ {\hat x}^\mu(\sigma_0) =x_0^\mu$ and there $ U^\mu {\hat S}_\mu =0$, we will regard ${\tilde x}(\sigma)$ to be a space-like curve orthogonal to $x^\mu(\tau)$. Again, ${\hat x}^\mu(\sigma)$ is another sort of  image of ${\tilde x}(\sigma)$.

   Similarly,  we may reverse the roles of $U$ and $S$ in the neighborhood  $\Delta{\cal V}^{ \prime}$ of any point $x^\mu(\sigma_0)$ of a space-like stream and define  local contiguous time-like  streams  $x^\mu(\tau)$ which are   orthogonal   to $x^\mu(\sigma)$ at $\sigma =\sigma_0$.

      One can consider  a  bundle of images  formed by replacing $A^\mu$ by a  small (weighted) bounded set  $\{A^\mu\}$  within  $\Delta{\cal V}$ as actually constituting the internal structure of  a stream (enclosing a central geodesic) itself.  Such a bundle of images is equivalent to representing its streamlines surrounding the central path $ {\tilde x}^\nu (\sigma)$.  This very local  representation is  useful in  interpreting solutions of the equations of motion.

        For the case when many space-like curves, labeled by the index $s$, become contiguous to the time-like stream, replace $\sigma, S^\mu,m_1, m_2$ with  $\sigma_s, S^\mu_s,m_{1s},m_{2s}$ and perform a summation over $s$. We suppose that this sum can also be replaced by a  integration  over frequency distributions as was suggested in the discussion of the summations used in defining $T^{\mu\nu}$.

 \subsubsubsection{ A6: The Vanishing Of The   `Line' Divergence  $U^\nu_{s;\  \nu}$}             %pg 26             
          The technical point here is that we are are considering functions defined only along a line so we cannot directly use the Gauss divergence theorem with no sources or sinks present.  However $U^\nu_{s; \nu}$ is  a scalar, independent of a coordinate system choice. At any point, choose a geodesic coordinate system and rotate the coordinate axes so that $\tau$ is the coordinate along one of the axes. Then $U^\mu$ has only component, $\mu=\tau$, and   $dU^\mu/d\tau =0$ because $U^\tau$ is  a  geodesic's tangent vector.
          
\subsubsubsection{A7: Close Sets of Streams.}  
    if locally all the streams are close, neighboring streams  may always have short-range interactions in ${\cal R}$ and the picture needs modification.  We defined streams to exist only between `collisions'. If streams can then begin and end within ${\cal R}$ , one should replace the $U^\mu_{s;\mu} =0$ term by say, the Boltzmann collision operator: this  preserves 4-momentum upon integration in the local phase space and preserves the fluid form of conservation of ${\bar N}$.] Then the frequency distributions of both the $U^\mu_s$ and the $\delta U^\mu_s$ are modified  and not really known until specifics of the collisions are specified.  A thermodynamic empirical form for  $P^{\mu\nu}$ would summarize this procedure in a particular coordinate system and again all four constraints would be needed.

\section{Appendix B:  Supplementary Physical   Arguments.}     
  
 \subsection{B1: Stream Momenta Conservation Conditions: } %see Feb 26, Mar 8, Mar 11 2016 for checks    %pg 27
 
  At a point  $x^\mu_0$ in a local Minkowski space ( within  $\Delta{\cal V}$) choose the particular Lorentz frame\footnote{ The condition $S^0=0$ is used to define $\Gamma, k$ uniquely.  If $K$ is tangent to a curve passing through $x^\mu_0$, then at neighboring points along this curve $S^0(x) \not= 0$ in general in this Lorentz frame.  Also, in another Lorentz frame $S^0 \not= 0$.  The introduction of $S^\mu$ -- with $S^0=0$ in {\it all} Lorentz frames -- was used in Weinberg (1960) to discuss classical systems with intrinsic spin.  This work helped inspire this approach.}
 in which $S^0=0$. Put  $S=(0,0, \sin \xi,\cos \xi)$ and $K=(\sinh \eta,0,0,\cosh \eta) \equiv \Gamma(k,0,0,1) $ for space-like vectors. 
                                                                                                                                              %pg. 12
     Then there is a one-parameter ($\sin \xi$) set of solutions for Eq(5) for which $S\cdot U=0.$  Set ${\bar a}U=(\sinh \eta, 0,-{\bar b}\sin \xi,
     {\bar b}\sin \xi \tan \xi)$ where ${\bar b}=\cosh \eta \cos \xi$, and ${\bar a}^2={\bar b}^2-1$.  If we introduce $u_z=U^z/U^0$ then
     
        $$ku_z =\sin^2 \xi;    \eqno(B1)    $$                                                   
        since for halo DM we expect $u^z \leq \ \sim 10^{-2}$, we need $\xi \ll 1$.  Then to high order, the specification of ${\bar b}\cong \cosh \eta$ and ${\bar a} \cong \sinh \eta$ is independent of the particular choice of $\xi$.\footnote{ Since we are exploring models for DM, whose properties are unknown, we should point out that $\xi$ and $k$ could be variable, {\it e.g.} specified by a path parameter $\tau$  associated with $U^\mu \equiv dx^\mu(\tau)/d\tau$.}  Since $U^z/U^y =\tan \xi \sim \xi$, the transverse velocity component $U^y$ is always dominant.  Then, a more useful parameterization is to  use $u_z$ itself and to express the transverse components  by
        $$(U^y) ^2 =u_z/k , \ \ (S^y)^2 =ku_z,      \eqno (B2a)$$
        and  (when $U^t \approx 1, \ U^y \cong u_y $) the translational components by
        $$U^z \cong u_z, \ \ S^z \cong 1-ku_z/2 ,    \eqno(B2b)$$
       with $(U^y/U^z)^2 \gg1$ and $(S^y/S^z)^2 \ll 1$.  Note $k$ is not required to be especially small.
  
    \subsubsubsection{Time-like  total momenta}
               If $K$ is time-like, set $K=(\cosh \eta,0,0,\sinh \eta) \equiv \gamma(1,0,0,v)$. Then, ${\bar a}U^0=\cosh \eta$; also ${\bar b}=\sinh \eta \cos \xi$ and ${\bar a}^2 ={\bar b}^2 +1$.   Then  the restriction $u_z =v \sin^2 \xi$ with  $u_z\ll 1$ again requires $\xi$ to be small  when $K$ is  relativistic;  so ${\bar a}= \gamma$ and $ \ {\bar b}=\gamma v$.   When $K$ represents non-relativistic ordinary matter (with  $v \ll 1$), $\xi$ may not be small;  then  ${\bar a} \cong 1$ and $ \ {\bar b} \leq v \ll 1$.    [For the null case when $K^\mu=(1,0,0,1)$ one finds ${\bar a}={\bar b} \cong 1$ with $u_z=\sin^2 \xi$. ]  
                
  \subsubsubsection {The form of the stream density}                       %pg 28
     Further, consider  two vectors orthogonal to $U,S$ such as:  $B^\mu=(0,1,0,0)$ and $ A^\mu =A^y(-\coth \eta \tan \xi,0,1,-\tan \xi)$. 
     There is a path through the origin $x^\mu(q)$ with  tangent  $dx^\mu(q)/dq =A^\mu$. Then the solution of the stream momentum density equation,  $dm_\tau/d\tau =U^\mu m_{,\mu}=0$ is
     $m_\tau= h_0(\sigma, x, q)$ where where $x^1=x$ and $h_0$ is arbitrary.  Similarly for $dm_\sigma/d\sigma =0$, one has $m_\sigma = h_1(\tau, x, q)$.  For  Eqn 4.1, set $m(x) =h(q,x)$.  We conclude that we may set $U^\mu_{;\mu}=0$ and  $S^\mu_{;\mu}=0$
     simultaneously and regard  $m_\sigma$ and $m_\tau$ as independent even if they have a common variable factor.

 \subsubsubsection{ The Value Of $u_z$. }  
                                % 6/9/16     
      There must exist a minimum value of $u_z$ for this representation  to be useful. 
      If something like standard cosmology holds, then the streams are required to equilibrate with the background cosmological fluid  through gravitational interactions  (and stream-stream interactions) until decoupling. Immediately after, using Eq(B2a) one finds
       $$\langle  u_y ^2\rangle +\langle u_z^2\rangle \simeq v_d^2, \eqno(B3)$$
     $$ u_z  \simeq kv_d^2 \ \ {\rm and} \ \ u_y^2 \simeq v_d^2 ;   \eqno(B4)$$
   where the characteristic decoupling velocity $v_d$   probably exceeds the value characteristic of  hydrogen when photons  decouple  $cv_d \geq 10\  {\rm km}\ {\rm s}^{-1}$.  
    
      \subsection{B2: The Guiding Center Solutions.}                                 % (see 6/21,22/16 for re-checks and history)  %pg28
   
      We adopt the guiding-center  approximation in making a simple local model, so that the only coordinates in the plane of spin are $(x,y)$. 
       Then  $U^x,U^y,S^x,S^y$ are functions of $x,y$ alone.   For the `coupling' function in  ${\cal L}_{US}$ take $m_2\equiv \surd (m_0m_1) f(x,y) \geq 0$, where $f$ is arbitrary.  For a local solution for the geodesics, Eqns (6a,b),  work  in cartesian coordinates in a Lorentz frame in which  $S^0=0$. In this local space  take $U^0, U^z,  S^z$ constant and define the   density factors as $m_0 ={\bar a}mj, \  m_1={\bar b}mj$  where $m=m(z,t)$ and $j=j(\varpi)$ is a `cut-off' factor.   Here $j$ is  centered on the traveling `origin'  $x^\mu_{00}(\tau)=(U^0 \tau, x_0,y_0, U^z \tau)$ and    equal to zero for $\varpi \equiv \surd[ (x-x_0)^2 +(y-y_0^2)] >  \varpi_{00}$.  First, let us take $x_0,y_0$ as locating the center of  $\Delta {\cal V}$. Then  $\Delta {\cal V}$ can be regarded as the small sheath surrounding the $z-$ axis, enclosing parts of both streams (the `helices').

      Recalling $\hat U =m_0 U$ and $\hat S=m_1S$,  multiply Eqn(6a) by $a^2$ and Eqn (6b) by $b^2$, where $a^2/b^2 =m_1/m_0$. [For halo DM,  $a^2/b^2 =\Gamma/(k\Gamma)$.] Then introducing the scaled velocities 
      $$(u^x,u^y)=a(\hat U^x,\hat U^y)\ {\rm and}\  (s^x,s^y)=b(\hat S^x,\hat S^y)         \eqno(B5a)$$    %pg 29
       one finds Eqns( 6a,b) reduce to the same equation form,
      $u^\beta u^\mu_{\ ; \beta} =g^{\mu \alpha} (fs)_{[\beta,\alpha]}u^\beta,$ with the interchange $ u^\mu  \leftrightarrow s^\mu $ giving the second equation. The two equations  can be reduced to one if 
       $$(s^x,s^y)=  -(u^x,u^y),    \eqno(B5b)$$
    imposing transverse momentum conservation.
      
         [The original normalization conditions become,
         
       $$ u^2=a^2m_0^2 [(U^0)^2-(U^z)^2-1] \equiv a^2 h_u^2(z,t),  \eqno(B6a)$$ 
       $$s^2= b^2m_1[1 -(S^z)^2] \equiv b^2 h_s^2 (z,t),  \eqno(B 6b)   $$
       
        where $ h_u^2(z,t)= h_s^2 (z,t)$  is required to conserve momentum. We will assume the spin is very rapid, so that in an inertial frame we may replace Eqns(B6,a,b) with $\langle \ dh_u^2(z,t)/d\tau\ \rangle =0$ when considering the variarions in  $u^\varpi, u^\varphi$ .  In effect, this guiding center approximation requires $\tau$ be regarded as split into two parameters, $\tau_0 +\delta \tau$, with $\delta  \tau$ determining the rapid `spin' behavior; similarly with $\sigma \to \sigma_0 + \delta \sigma$.]

        In local cylindrical coordinates, Eqn (B5) becomes:
        $$u^\varpi u^\varpi_{,\varpi} +u^\varphi [u^\varpi _{, \varphi} -\varpi u^\varphi -(\varpi^2 fs^\varphi)_{,\varpi} + (fs^\varpi)_{,\varphi}] =0;   \eqno(B5a)  $$
        
       $$\varpi^2 u^\varphi u^\varphi_{,\varphi} + u^\varpi [\varpi^2 u^\varphi_{,\varpi}+2\varpi u^\varphi -(fs^\varpi)_{,\varphi} +(\varpi^2 f s^\varphi)_{,\varpi}]=0.  \eqno(B5b) $$
       (where now we have replaced $f/2$ by $f$ for notational convenience.)  One also has the corresponding equations when the exchange  $ u^\mu  \leftrightarrow s^\mu $ is made.

        A simple  solution for the averages  satisfying Eqns(B5a,b) when $f=f(\varpi)$ is obtained by setting $u^\varpi =0$ and $s^\varphi=u^\varphi$
        and requiring 
        $$ \ln (\varpi^2 f u^\varphi) = -\int d\varpi/(\varpi f ) .       \eqno(B8)$$ 
                                                                                                                                        
        For example, using $f=(1 + \sigma \varpi)^{-1} $ to define the extent of  $\Delta {\cal V}$, one finds  $\varpi^3u^\varphi =k_u(1+ \sigma\varpi) e^{-\sigma\varpi}$, with $k_u$ a constant.  [Similarly   for the  $s^\varpi, s^\varphi $ equations  with  $s^\varpi=0$, one gets the same result with $k_u$ replaced by  $k_s$.] The normalizations, Eqns(6a,b),  {\it e.g.} $\varpi_s^2(u^\varphi)^2 =h_s^2(z,t)$, determine the radii and the `constants' $k_u,k_s$, once $u^\varphi$ is specified.  As $h_s^2, h_u^2$  slowly change with position, the radii $\varpi_u, \varpi_s$ also change. 
        
             The additional assumption that the total transverse momentum of the pair of streams is conserved is imposed by setting   $\varphi_s =\varphi_u + \pi.$                                         %pg30
            
        Suppose $u^\varphi,\varpi$ are taken to be constant. Then Eqn(B8)  determines a  relation that these local constants must  obey. Along the stream's path $u^\varphi, \varpi$ may slowly change from one constant set of values to another, but both sets should satisfy Eqn(B8) to preserve the total angular momentum. 
        
        The coupling of the $u-$equations and the $s-$equations  is through the velocities.  Suppose we change the origin $(x_0,y_0)$ to the center of the $u-$stream; the equations do not change. We may then regard Eqn(B8) as giving the variation of $u^\varphi$  across the stream. For this choice of $f$, one sees $2\pi \int  (\varpi^2 u^\varphi) \varpi d \varpi \to {\rm const.\  for} \ \sigma \varpi \gg 1$;  the total angular momentum carried by the stream is constant.       
      
\subsubsubsection{An Interpretation Of  The Streams' Motions.}

    Eqns (B5a,b) each hold on the different paths  $x^\mu(\tau), \ x^\mu(\sigma)$  inside $\Delta {\cal V}$. To solve, we have effectively 
    made two transformations into a common path resulting in `scaled' solutions:   $am_0x^\mu(\tau) \to \xi^\mu (\tau_0 + \delta \tau)$ and $bm_1x^\mu(\sigma) \to -\ \xi^\mu (\sigma=\sigma_0 +\delta \sigma)$ in cartesian coordinates. The two streams actually are braided, like DNA, with different radii for the two helices; together they act like a spinning system translating in the $z-$direction.       These are called `scaled' solutions for the following reason:
 Take a particular point $(x_0,y_0)$  on $x^\mu(\tau)$ for evaluating ${\hat U}$ so that $am_0  x_0 =\xi_0 ,am_0 y_0=\eta_0$. It corresponds  to a point $(x_1,y_1)= -(b/a)(x_0,y_0)$  on the $x^\mu(\sigma)$ path;  corresponding points in the two streams `opposite' each other with respect to  the $z-$axis are  brought  to the same image point in  $(\xi,\eta)$ coordinates.

\subsubsubsection{ The Local `B'-field, Collisions \& Assignment Of The Spin Velocity.}

            From the Lagrangian formalism, a particle physicist would normally regard the term ${\cal L}_{US}$ as describing a short range interaction  with $f(x)$ determining the local `force' between  two `fields'. This interpretation is not  germane  to this model. But the magnitude of $f$ determines the spin rate within the two streams, {\it i.e.} the internal velocity structure of the braided streams, and $f$ may not be the same for all DM braided streams. 
            
            Also since  $2 \oint (fS)_{[x,y]}\varpi d \varpi =2\pi \varpi^2 B_{eff}$, by Stokes' theorem, a $B-$like {\it field}, dependent upon $f$, is produced around the streams.  We  expect other pairs of streams entering $\Delta {\cal V} $ would be able to interact with the braided pair through this  field. We suggest this interaction plays the role of a `collision' allowing neighboring streams to modify the magnitude of $f$, and hence of $u^\varphi$.  Consequently, for  an ensemble of DM streams in equilibrium  we shall assume the mean spin velocity $u^\varphi$ is  not fixed but has the value of  the r.m.s. of the peculiar velocity distribution. We noted that the rotational energy did not enter $T^{\mu\nu}$;  we expect a more sophisticated argument would  regard it contributing to the tensor $P^{\mu\nu}$ (see Eqn(3a)) which we have disregarded.
            
\section{  B3: Halo Kinematics.}                                                                                                                 %pg31

     In the Kepler central force problem, one needs to specify the energy, angular velocity and initial position of a moving body.  It turns out that we require the same information for a representative DM stream in order to specify the false pressure and density structure within a halo.

 \subsection{ Representation Of The Averaged $T^{\mu\nu}$ In The DM Halo Model.}          
       In this  very simple halo model we assume no OM and only one type of DM, that for which the energetics at $r \geq r_q$ , intergalactic space, are given by $K: \Gamma(k,0,0,1)$ in a local Cartesian frame ; See Eqns(5).  The DM  follows orbits satisfying the integrals of motion; see Eqns(9a,b,c,d).  We choose  $r =r_p$ to be the turning point of  a representative stream  $U^r (r_p)=0;$  at that radius,  $(S^r(r_p))^2 >  0$ (See  Section 4.0.1).  For $r \leq r_p$ the solution for $S^\mu$ has no physical significance and one may take $T^{\mu \nu}=0$.

         In the annular region $r_p \le r  \leq r_q$, one specifies the halo $T^{\mu\nu}$ by the change of the halo $T^{\mu\nu}$ from its value in the background model at  $r= r_q$ caused by the compression of the pressure in this inner region.   We may write  for the change in a typical  spacial diagonal component  of $\delta T^{\mu\nu}$ at the  halo point  $r\geq r_p$,  
      
      $$\delta T^{xx} = {\bar m}\  {\cal N} {\bar P} - {\bar m}_q\  {\cal N}_ q {\bar P}_q,    \eqno(17)$$   
      
       expressing the halo pressure increment at $r$.  Here ${\bar m}$ is an average stream density, ${\cal N}$ is the number of streams in a unit volume and ${\bar P}$ represents an average value of the square of the `speed'  $s^2\equiv (S^r)^2   +  J_s^2/r^2 $ along the paths $x^\mu_s (\sigma_s)$;  it corresponds to $ \langle(S^z)^2\rangle/3$  of Eqn(15). A similar expression holds for $\delta T^{00}$ in terms of the average value of $(U^0)^2$. 
       
                 Further we restrict ourselves to a region $r \leq  r_a < r_q$.  We will assume a galaxy going through the intergalactic medium  acquires a stand-off bow shock and between this shock and the galaxy there is a  zone of compression and stagnation, occurring in this simple  model at $r=r_a$. Material from this zone falls inward at nearly parabolic speeds, $v(r) \simeq \surd (2 \Psi) \approx \surd (2G{\cal M}_{eff}/r) $.

       Then, using the steady state one dimensional pipe flow, one has $ {\bar m}_r / {\bar m}_a = s_a/s_r$.  Also,
       ${\cal N}_ r/{\cal N}_ a =  r_a^2/r^2 $ in a free-fall zone and ${\cal N}_ r/{\cal N}_ a =  v^2(r)/v_a^2$ in an inner zone in which conservation of angular momentum impedes radial motion. ( See Appendix B3.)  Finally, from Eqn (13),
      $ {\bar P}_r -2\Psi_r ={\bar P}_a-2\Psi_a$ where $\Psi$ is the potential associated with the metric element $B(r)$. (See Section 4.)  Assuming no viscous coupling between the DM streams, an expression for the pressure variation $\delta p$  in the halo is given by Eqns(B9a,b).  A detailed model of the halo can be now constructed; it is given in Appendix C.

\subsection { B3: Properties Of  Collections Of In-Falling Halo  Streams.}       % pg32    (for numerical examples see 2/6,7/17.)
        From Eqn(15) one has      
     $$\delta p =m{\cal N}P -m_a{\cal N}_a P_a= \left[{m\over m_a}{{\cal N}\over {\cal N}_a}{ P\over P_a} -1\right] p_{aa},  $$
     
     where $P\equiv \langle (S^z)^2 \rangle /3$ was discussed in Section 3.2, (see Eqn 9) and we have set $p_{aa} \equiv m_a\ {\cal N}_aP_a$.  Since $S^z$ is  very large compared to $\Psi$, we may take it not very much affected by the gravitational field in the halo and we can set 
     $m/m_a =S^z_a/S^z \simeq 1 -\Psi_a+\Psi \sim1 \ \ {\rm and}\ \ P/Pa =(S^z)^2 /(S_a^z)^2 \simeq 1-2\Psi+2\Psi_a \sim 1$. Hence for DM
    $$p/p_a ={\cal N}/{\cal N}_a.$$
         In a halo in-falling streams pass through two zones: a `free-fall' zone and a zone in which radial motion is severely limited by conservation of angular velocity.  The boundary between the two
         is a sensitive function of the initial angular momentum.
         
         We can introduce a statistical treatment of large numbers of  time-like streams (with $U^0_s>0$)  similar to  the treatment of ``pencils of radiation" in  standard radiative transfer theory.\footnote{ See Chandrasekhar, 1950; Rybicki \&  Lightman 1979.}    Normally one writes for the number of streams  crossing a surface element $dA$ in a solid angle $d\Omega$ oriented at an angle $\theta$ to the normal of $dA$ in a time interval $dt$,
       $$dF=(I/\pi) \cos \theta \  d\Omega dA dt, $$             
               defining $I$.  For a steady-state when $I$ is independent of $\theta,\phi$,  the total number of streams crossing $dA$ in one direction  in unit time is $dF=  I dA$. [ This corresponds to the particle kinetic theory result ${\hat F} ={1\over 4} n v$ for the flux of particles crossing a unit area.]  We see $I \propto {\cal N}$ of Eqn(15), the local density of streams.          %pg 33
              
          Consider only those area elements whose normals are the radius vectors.    For convenience suppose  at some large $r = r_a$,  in a volume element all the streams have the same density $\langle m_a\rangle$ and the same average speed $v_a$   (where $U_s^\mu =\gamma(1, \vec v_{a,s})$).We calculate the usual density  
            $\rho=\langle m_a\rangle {\cal N}$ by first calculating the $r-$dependence of ${\cal N}$ (assuming space is flat).                
            
  For each stream the orbital angular velocity $L_s=r_a \times (r_a U^\phi_s) = v_a  \ r_a  \sin \theta_s \equiv v_a b_{a,s} $ is  a constant and is determined just by the value of   $\theta_s $ ( the angle between the 3-vectors $r_a$ and $\vec v_{a,s}$). The total number of {\it inflowing} streams  at  a shell of radius $r_a$  is then $ F_a={1\over 2}\cdot 4\pi r_a^2 \cdot \int I_a \cos \theta \ \psi(\theta, \phi) \sin \theta d\theta d\phi$, where $\psi$ is a distribution function for the ingoing streams.  For large $r$ we adopt a distribution $\psi=1$, corresponding to velocities being isotropic\footnote{The specification of $\psi$ is part of the specification  of the angular velocity distribution; in principal it could be a function of $v_ar_a$.  In near equilibrium the  symmetries of the  velocity distribution reflect the symmetries of the effective gravitational potential;  see  Chandrasekhar (1942).} so $ F_a ={1 \over 4} \cdot 4 \pi r_a^2 I_a$; these streams carry orbital angular momenta up to $L_a=r_a v_a$.  
  
    Let us follow such an aggregate falling in towards the center, assuming for convenience that $v_a$ is small enough that each stream follows a nearly parabolic orbit, {\it i.e.} $v_a^2 \ll  V_0^2(R_0)$, where $V_0$ is close to the galaxy's maximum rotation speed. For large $r=\tilde r< r_a $, a similar argument gives ${\tilde F} =\pi \tilde r^2\tilde I$ for the  inflowing streams.  There is a region of `free-fall' in which the streams are not appreciably deflected;  in it  $\tilde F =F_a$ so 
    
    $$\tilde I/I_a ={\tilde {\cal N}}/{\cal N}_a =  r_a^2/{\tilde r}^2  =\tilde p/p_a.     \eqno(B9a)  $$   
    
       Inside this region streams with large  orbital angular velocity $L_s$ will be deflected. We calculate the loss to the inflow.  In the background space suppose we regard each sphere $r= {\rm const.}$ as a collection of  `bound' DM circular orbits, with an associated orbital angular velocity given by $L_{cir}^2=r^2 v_{cir}^2(r)$.  For a falling stream to pass through a sphere of radius $r$ it must have 
  its distance of closest approach to the center be less than  $r$.  For example, let $\Psi =G{\cal M}/r$. For DM following a parabolic orbit one requires   $L^2_s \leq 2L_{cir}^2  +W_u^2 \equiv L_\ast^2(r) ;$   define    $r_p \equiv W^2_u/(G{\cal M})$; then $L_{cir }^2 =G{\cal M}[r_{cir} -r_p]$.        
  
                 % for orbit details see final check 3/2/17 pg 1.
       
       Consider a particular sphere $ r=r_b.$ Only those streams with low $L_s$ reach $r_b$; they have the same (invariant) values of $L_s$ as they had when they were at $r=r_a$.  We calculate this fraction $f$ of inflowing streams at $r_a$ by using  $L=r_a v_a \sin \theta$ as the variable of integration instead of $\theta$; one has $f =(r_a v_a)^{-2}\int_0^{L_\ast} LdL $ or
       
       $$ f \cong r(r-r_p/2 ) V^2(r)/(r_a^2v_a^2), \eqno (B10a)$$   %pg 34
       
where $r_p$  is the effective  inner  halo edge, defined by the intrinsic angular velocity, and we  have put $V^2(r)\equiv G{\cal M}/ r $. [Note: parabolic infall velocities have their turning point at $r=r_p/2$ and `bound'
circular orbits cannot be defined for $r \leq r_p.$]  
    
        This fraction must be compensated for by an increase in $I_b$ for the undeflected streams since the total number of streams  which can travel from $r_a$ to $r_b$ is conserved.   Consequently\footnote{ Because we have not assumed that at our starting point, $r=r_a$, $v_ar_a$ is the maximum orbital angular velocity possible, $v_a^2 \leq v_{cir}^2(r=r_a)$.} since $\pi r_b^2I_b = f \cdot \pi r_a^2I_a$, one has, for an inner zone
       
       $$I_b/I_a ={\cal N}_b/{\cal N}_a =  p_b/p_a =\tilde v_b^2/ v_a^2 \equiv V^2(r_b) [1-r_p/(2r_b)]/v_a^2,   \eqno(B9b)$$     
       
defining $\tilde v(r_b)$;      this holds\footnote{ We have assumed that in the outer halo $r v_{cir}$ for DM is an increasing function of $r$, so that there is a limiting radius $r_b^\ast $ at which `free-fall' ends and for which $r<r_b^\ast$  Eqn(B9b) holds.}  for all $r_b \leq r_b^\star$ for which $  \tilde v_b r_b \leq  r_a v_a$. 
 The boundary between the two regions, $f=1$ occurs at $r=r_b^\star$ where
 
 $$ r_b^\star -{r_p \over 2}=v_a^2 r_a^2/(G{\cal M}_0 ) \eqno(B10b)$$
  
  assuming $r_b^\star$  is close to the inner halo edge  where  $G{\cal M}(r) \simeq G{\cal M}_0 \ (=r_b^\star V^2(r^\star_b) = R_hV^2(R_h))$.  This is true for high mass galaxies and low initial angular velocities, $r_av_a$;  this is the case we emphasize. If this holds the DM  generally is in the free-fall zone where $p \propto  r^{-2}$. But, for low-mass galaxies $r_b^\star$ may be close to $r_a$,  severely limiting the `free-fall' zone.

  \subsubsubsection{  The Halo `Density' \& `Pressure' As Functions of $r$,}   
  
  We study the region $r \leq r_a.$  Choose as unit of length  $R_h =r_a^2v_a^2 /(G{\cal M}_0)$. Represent the intrinsic angular velocity by
  $W_u^2= r_a^2 (\xi v_a)^2$ with the typical orbital angular velocity given by $L_u^2 =r_a^2v_a^2$. Then the formal end of the halo, defined in terms of the lowest  bound `circular'
 path is $r_p =\xi^2 R_h$ and the furthest inward DM parabolic orbits can go is $r_p/2 \equiv \eta R_h$;  we will use $\eta \approx {1\over 2}$  as representative in the approximations because we are interested only in infall with low angular momentum.  In these units  the boundary between the outer region and the inner region is $r_b^\star = (1+\eta)R_h$.    
 
      In summary, the halo structure starts at $r \cong R_h$.  The false pressure and density are nearly constant out to $r =r_b^\star$ and then fall $\propto 1/r^2$ as we go further out.        %pg 35
 
     In the outer region one has:
 $$ p(r)=p_a (r_a/r)^2;\ \ \rho(r) / \rho_a =[\Psi (r_a)/ \Psi(r)]^{0.5}  p(r)/p_a \  \ r_a \geq r \geq r_b^\star.  \eqno (B11a),$$ 
 
  To lowest order we ignore the variation of $\Psi$ in the outer zone and use  $\rho(r) / \rho_a \approx p(r)/p_a $. In the inner zone, using $p_b =p_a (r_a/r_b^\star)^2; \ \rho_b =\rho_a(r_a/r_b^\star)^2$ and $r=xR_0$ (so that $\eta \leq x \leq 1+\eta$),          % approximations checked 4/24/17,4/25/17
 
 $$ p  =(1+\eta)^2p_b (x^{-1} -\eta x^{-2} ) ; \ \ \rho=(1+\eta)^2 \rho_b [ (x/(1+\eta)]^{0.5}(x^{-1} -\eta x^{-2} ). \eqno(B11b)$$
For the DM mass in the inner zone, using $M_1 =(4\pi/3)\rho_bR_0^3$, one has
 
 $${\cal M}_1(r) =M_1(1+\eta )^{3/2} \ [ 6x^{5/2}-10\eta x^{3/2}  +4\eta^{5/2}]/5, \eqno(B11c) $$
 Also, for  the additional DM mass above the boundary, $r_b^\star$, in the outer zone, one finds
 
 $${\cal M}_2(r) = 3(1+\eta)^2 M_1[x-(1+\eta)] . \eqno(B11d)$$
 With these expressions for $p,\rho$ the structure of the halo can be developed. See Appendix C.
 
 \subsubsubsection{ Approximations And Limitations. }
  One may use  $p \approx  p_b \ ,\ \rho \approx 0.8\rho_b\surd x \sim \rho_b  \ {\rm for} \ 1\leq x \leq  3/2 $ and $p,\rho \sim 0 \ {\rm for}\  x < 1$ (using $\eta =1/2$) since the inner region, $1+\eta \geq x\geq 1$, is a zone of  (nearly) constant density and pressure.       For estimating mass contributions it is adequate to use the  linear approximation 
${\cal M}_1(r) \approx 5 M_1(x-0.9)  \ {\rm for}\ 0.9 \leq x \leq 1.5 $  and   ${\cal M}_2(r) \sim 7 M_1 [r/R_h]$. Also  ${\cal M}_1(r_b^\star) \cong 4\pi\rho_b R_h^3$.  For the Newtonian potential one has $\Phi(r) =G[({\cal M}_0 + \delta {\cal M}(r)]/r $, where $\delta {\cal M} ={\cal M}_1(r)$ in the inner zone and $\delta {\cal M} ={\cal M}_1(r_b^\star)+{\cal M}_2(r)$ in the outer zone.   

   We expect that the formulae in the inner zone may depart significantly from those for the spherical point source model used here if $R_h$ is small ($ \sim 5$ kpc). Then the potential of the central galaxy may depart significantly from the spherical point source model used. They   are useful for contrasting the DM infall  from that of OM when $\eta =0$; in that case {\it e.g.} ${\cal M}_1 \propto x^{5/2}$ and $v_{cir} \propto  x^{3/4}$, assuming no particle interactions.   
     
       The boundary $r_b^\star$ location is very sensitive to the initial orbital angular velocity assumed.  If the mean speed at $r=r_a$ is small compared  to $V(R_h) \equiv \surd (G{\cal M}_0/R_h)$,  then $r_b^\star $ may be  close to $r_p/2 $.  For example, if $v_a  \leq 20 \ {\rm km\ s}^{-1}$ and $V(R_h) \sim 200   \ {\rm km\ s}^{-1}$ and one chooses  $r_a   \sim 50 \  {\rm kpc} \simeq 5-10 \ R_h  $  then   $r_b^\star -r_p / 2 \leq 2   \  {\rm kpc}$ so that the inner zone effectively is $\leq \sim 1$ kpc in extent.  
  \subsubsubsection{The Maximum Value of $r$ In Inflows.}      %pg 36
   To avoid an Oblers paradox situation from developing, even in these simple inflow models a maximum value  $r_a$ must be recognized. 
   We suggest it results   from the velocity  of the central galaxy being supersonic with respect to the cold dark intergalactic medium. In the rest frame of the galaxy, a bow-shock discontinuity and tail must arise, enclosing a stagnation volume in which the flow  is subsonic. In our simplified model we shall take the stagnation volume as spherical  with  boundary at  $r =r_a$. We assume the  large external velocity is converted to an  interior flow with a much smaller subsonic velocity $\hat v_a$;  then the conservation laws across the   transition zone must include ${\cal N}_{ext}V_{ext}  =  {\cal N}_a \hat v_a $; similarly for $m{\cal N}$. [For convenience we also assume  that the translational velocity $\hat v_a$ is  small compared to $v_a$, the r.m.s. velocity used in establishing Eqn.(B9) so that the calculation of the transport of orbital angular momentum, resulting in Eqn (B11), is simplified.]  
   
   At such a boundary the values of $\rho_a,\ p_a$ would be enhanced over their intergalactic values by the factor  ${\cal N}_a/ {\cal N}_{ext}$; using  $V_{ext} \sim \ 200 \ {\rm km \ s}^{-1} $ and $ \hat v_a \sim \ 2-20 \  {\rm km \ s}^{-1} $ as reasonable estimates one sees an enhancement factor of  $\sim 10\ - \ 100$ may occur.  We suggest, as a very rough estimate that
    $r_a \sim 5-10\ R_h \sim 50-80\ {\rm kps}$.
    
    There is another approach for determining $r_a$.  There is a  theoretical `shielding' distance  $R_G$ such that 
    $4\pi G\rho_a r_a^2 +G{\cal M}_0/R_G=0$ (since $\rho <0$).  Only for $r\leq R_G$ does the central galaxy exhibit an attractive force; consequently $r_a \leq R_G$.

\subsection{ Appendix C: Details Of A Halo Model For An Isolated Galaxy.}    %pgs 37

  The DM halo is a pressure dominated structure. It has three characteristic distances: (1) an inner edge,   $r_p/2$, determined by  the centrifugal barrier  to inner directed motion of the DM braided streams; (2) $r_a$, the edge of the region of significant pressure compression, $r_p \leq r \leq r_a$; (3) an outer `edge' $r_q$,  beyond which the gravitational attraction of the central object  is not important. 
  
      The  compressed region  $r_p \leq r \leq r_a$ is divided  into two zones; the inner zone $r_p \equiv 2\eta R_h \leq r \leq r_b^\star \equiv R_h(1+\eta)$ in which radial motion is strongly impeded by centrifugal forces; and the outer   `free-fall' zone $r_b^\star \leq r \leq r_a$ , in which radial motion is unimpeded.  In the very outer uncompressed halo region, $r_a \leq  r \leq r_q$ , the density $\rho_{00} $ and pressure $p_{00}$ (where $ p_{00}/c^2 \  \gg \rho_{00}$) match that of the DM intergalactic medium of which it is a part; in  theory these are  provided as boundary conditions on the model  but in practice the the two halo potentials $\Phi,\Psi$ have their constants adjusted to match the potentials for the central  galaxy, so that  $p_{ 00}, \rho_{00}$ should be determined from observations. Here the transition from the halo model to that representing intergalactic space is arbitrary so that $r_q-r_a$ can be small. 
      
      We use something in the form of the standard fluid energy momentum tensor $T^{\mu\nu}= (p+\rho)U^\mu U^\nu -pg^{\mu\nu}$  (with $c^2=1$) as the halo field source term;  the special values of $\rho,p$  used  for DM  were discussed in Sec. 5.1; one has $p$, the `false' pressure term, as large and $\rho <0$ as comparatively small.
 
    \subsubsubsection{Outside}                                                  %pg37
  The intergalactic medium is represented by the Robertson-Walker metric
  $$d\tau^2 =d \bar t^2 -R(\bar t)[ (1-\bar kr^2)^{-1} dr^2 +r^2d\theta^2 +r^2 \sin^2 \theta d\phi^2]    \eqno(C1a)$$
   where the structure equations
   $$ 3\ddot{ R} =-4\pi G(\rho + 3p)R, \ \ \dot{ R}^2 +\bar k =8\pi G (\rho R^2)/3              \eqno(C1b)$$
   and  the `energy' conservation constraint
   $$3(p+\rho) d \ln R/ d \bar t =-d\rho /d \bar t        \eqno(C1c)$$
    holds for $r>r_q$. One can get a static model  (with $R=1$) by adding to the source  a  constant term, $-\hat \lambda g^{\mu\nu}$, so that $p \to \hat p= p +\hat \lambda, \ \rho \to \hat \rho= \rho -\hat \lambda$  allows one to set $\rho + 3p \to \hat \rho +3  \hat p  +2\hat \lambda = 0$.  This addition of a constant term is always admissible (See Sec. 2.)  We do not require that $p$ represents a `true' pressure nor that $\rho >0.$

\subsubsubsection {The Form Of $A(r)$ In The Halo.}      %pg 37/38 see 4/13/17 for final check

    In the Schwarzschild model (Eqns (10a,b))  for the entire halo $r_p/2 \le r \le r_q$, we have $A=1/(1-2G{\cal M}(r) /r)\equiv1/(1-2\Phi)$ where ${\cal M} ={\cal M}_0 +\delta {\cal M}$; ${\cal M}_0 $ is the mass of the central attractor  at $r \approx 0$ and $ \delta {\cal M}$ is given by Eqns (B11c,d); For DM , $ \delta {\cal M}<0.$  So, for example, $\Phi(r) =G {\cal M}_0 /r \ \
      {\rm for}\ \  r  \leq R_0 $, and using the approximations of Appendix B3,   one has
   $$\Phi (r) =G {\cal M}_0 /r +  4\pi G \rho_b r^2 (1- R_h^3/(r^3 ) /3 ,\ {\rm for}\ \  R_h \leq r \leq r_b^\star,   \eqno (C1) $$
       where $r_b^\star =R_h(1+\eta)$.  There is a similar expression for $\Phi (r)$ for $ r_b^\star \leq r \leq r_a$, resulting in {\it e.g.}
        $$ \Phi(r_a) = G( {\cal M}_0+{\cal M}_1)/r_a+ 4\pi G\rho_ar_a^2[1-r_b^\star/r_a],    \eqno(C2)$$
    where $\rho_ar_a^2 =\rho_b (r_b^\star)^2$ has been used.  One has
    
    $v_a^2r_a^2 =G{\cal M}_0 R_h$ and, for massive galaxies $[1-r_b^\star/r_a] \approx 1$.  [See Eqn(B10b).]          
\subsubsubsection{The Form of $B(r)$  In The  Halo}         %p38/39

     For the Schwarzschild metric,  we have $A^\prime /A+B^\prime/B =8\pi G(p +\rho)r A$. Again using $B=1-2\Psi$, where $\Psi$ is small in a halo model, one may rewrite this in a more useful form:
     $$v_{cir}^2 \equiv -r d\Psi/dr = 4\pi Gpr^2 + \Phi(r),  \eqno(C3)  $$
     where, for DM one has $ \delta {\cal M} <0.$
      Both $p$ and $\rho$ are continuous at the joining points $r=r_p/2$ and $r=r_b^\star$.  Using  the approximations of Appendix B3, for $ r_p/2 \leq r \leq R_h$, one has $\Psi(r) \cong \Phi(r)=G {\cal M}_0 /r$. In the  the interior halo zone, one has
     $$ v_{cir}^2 = 4 \pi G p_br^2 +G {\cal M}_0 /r +4\pi G \rho_b r^2[1-R_h^3/r^3]/3\ \ {\rm for}\ R_h \leq r < r_b^\star. \eqno(C3a).$$
     For the outer zone, since $p \propto 1/r^2$, one has
     $$ v_{cir}^2 = 4 \pi G p_ar_a^2  +4\pi G \rho_a r_a^2[1-r_b^\star/r] +G [{\cal M}_0+{\cal M}_1] /r  \eqno( C3b) $$
    for $ r_b^\star \leq r < r_a$.  Again $\rho_a r_a^2 \cong \rho_b(r_b^\star)^2$ , and $\rho_a, {\cal M}_1 < 0$. For large $r$, one has
    $v_{cir} \cong $ constant.
    
  \subsubsubsection{ The Outer Rim of the Halo.}    
    
             In a small  outermost halo region, $ r_q\geq r>r_a$   where $r_a$ is large and $\rho = \rho_{00},\ p  =p_{00}$ one has
             $$ \Phi(r) =   {\cal M}_T/r +4\pi G\rho_{00} r^2 /3 ,   \eqno(C4a) $$ 
             where $ {\cal M}_T=  {\cal M}_0 +\int_{R_h}^{r_a} 4\pi \rho(r) r^2 dr -4\pi \rho_{00}r_a^3/3 $ .  Introduce, for satisfying boundary conditions,
             $T^{\mu\nu}_{BC} = - \lambda g^{\mu\nu}$    so that     $ \rho \to \rho_{00} -\lambda, \ p \to p_{00} + \lambda$ where 
              $\lambda$   is chosen so that ${\cal M}_T \to 0$;  then $A$ is reduced to the form encountered in the Robertson-Walker metric.   Since $\Psi^\prime =\Phi^\prime +4\pi G (p_{oo} + \rho_{00})r^2$,  one has 
              $$v_{cir}^2 = 4\pi G [p_{00}+\rho_{00} +2(\lambda -\rho_{00})/3] r^2,    \eqno(C3c)$$
              which gives the familiar linear expansion (or contraction) with distance, encountered in cosmological problems.  Also when
              the annulus  $ r_q\geq r>r_a$ is very narrow,we  have
                            $$  Bdt^2 \equiv (1-2\Psi(r))dt^2 \to \cong(1-2\Psi(r_a))dt^2 \equiv d\bar t^2 ,  \eqno(C4) $$
                            which  again matches  the   Robinson-Walker metric form. So there is no problem in switching into the usual representation of the intergalactic medium. The equation also  expresses   the difference in clock rates between the   that of the intergalactic medium and that of the halo which is in a potential well.

    \section{References.}
    
     [1] V.C. Rubin, {\it Ap.J.} {\bf 238}, 471 (1980)
    
     [2] V.C. Rubin, {\it Science} {\bf 220}, 1339 (1983)
    
     [3] M. Fich and S. Tremaine,{\it ARAA} {\bf 29}, 409 (1991)
    
     [4] Y. Sofue  and V.C. Rubin, {\it ARAA} {\bf 39}, 137 (2001)
    
     [5] A.S. Kulessa and D. Lynden-Bell, {\it MNRAS} {\bf 255}, 105 (1992)
    
     [6] T. de Zeeuw and M. Franz, {\it ARAA} {\bf 29}, 239 (1991)
    
     [7] P.J.E. Peebles and B. Rata, {\it RMP} {\bf 75}, 559 (2003) 
    
     [8]  M. Milgrom, {\it Ap.J} {\bf 270}, 384 (1983)
    
     [9] H.L. Helfer in {\it Progress in Dark Matter Research; ed. J. Val Blaine} (Nova Science:NY) (2005)                    
                                                  
    [10] S. Weinberg {\it Gravitation and Cosmology} (Wiley:NY) 1772
     
    [11] E.W.Kolb and M.S. Turner, {\it The Early Universe} (Addison Wesley: NY) (1990) % esp. pg 48
   
    [12] S. Chandrasekhar, {\it Principles of Stellar Dynamics}(U. of Chicago:Chicago)(1942)
    
    [13] G. Battaglia {\it et. al} {\it MNRAS} {\bf 364}, 433 (2005)
    
    [14] D. Mihalas and P.M Routly , {\it Galactic Astronomy}(1st ed. W.H. Freeman:San Francisco) (1968)
   
    [15] D. Mihalas and J.Binney , {\it Galactic Astronomy}(2nd ed. W.H. Freeman:San Francisco) (1981)  
    
    [16] M. Tegmark {\it et al}, arXiv: astro-ph/0310723v2 (2004)
    
    [17] D.P. Clemens,{\it Ap.J.}{\bf 295},442 (1985)
    
    [18] V. Trimble in {\it Allen's Astrophysical Quantities}(4th ed.,edited byA.N.Cox Springer:NY) (2000)
    
    [19] J.N. Bahcall, M. Schmidt, and R.M. Soneira, {\it Ap.J} {\bf 265}, 730 (1983)
    
    [20] J.N.Bahcall, {\it ARAA} {\bf 24} 577 (1986)
    
    [21] G. Gilmore, {\it MNRAS} {\bf 207}, 223 (1984)
    
    [22] R. Mandelbaum {\it et al} {\it MNRAS} {\bf 368} 715 (2006)
    
    [23] N.A. Bahcall in {\it Allen's Astrophysical Quantities}(4th ed.,edited byA.N.Cox Springer:NY) (2000)
   
   [24] D. Scott,{\it et al} in  {\it Allen's Astrophysical Quantities}(4th ed.,edited byA.N.Cox Springer:NY) (2000)
    
    [25] R. Mainini,L.P.L.Colombo, and S.A. Bonometto {\it Ap.J.}{\bf 632},691 (2005)
    
    [26] L.D.Landau, and E.M. Lifshitz {\it Classical Theory Of Fields}(3rd ed. Pergamon:Oxford) (1971)

     \end{document}